\documentclass[manuscript]{aastex} \usepackage{emulateapj5}
\begin{document}

\title{Mass-Radius Relationships for Solid Exoplanets}

\author{S. Seager\altaffilmark{1,2},
M. Kuchner\altaffilmark{3},
C.~A. Hier-Majumder\altaffilmark{1},
B. Militzer\altaffilmark{4}}

\altaffiltext{1}{Department of Terrestrial Magnetism, Carnegie Institution of Washington, 5241 Broad Branch Rd. NW, Washington, DC 20015}

\altaffiltext{2}{Department of Earth, Atmospheric, and Planetary Sciences, Department of Physics, Massachusetts Institute of Technology, 54-1626, 77 Massachusetts Ave., Cambridge, MA, 01742}

\altaffiltext{3}{Exoplanets and Stellar Astrophysics Laboratory, Code 667, NASA/Goddard Space Flight Center, Greenbelt, MD 20771}

\altaffiltext{4}{Geophysical Laboratory, Carnegie Institution of Washington, 5251 Broad Branch Rd. NW, Washington, DC 20015}

\begin{abstract}

We use new interior models of cold planets to investigate the
mass-radius relationships of solid exoplanets, considering planets
made primarily of iron, silicates, water, and carbon compounds.  We
find that the mass-radius relationships for cold terrestrial-mass
planets of all compositions we considered follow a generic functional
form that is not a simple power law: $\log_{10} R_s = k_1 + 1/3
\log_{10}(M_s) - k_2 M_s^{k_3}$ for up to $M_p \approx 20 M_{\oplus}$,
where $M_s$ and $R_s$ are scaled mass and radius values. 
This functional form arises because the common building blocks of
solid planets all have equations of state that are well approximated by a
modified polytrope of the form $\rho = \rho_0 + c P^n$.

We find that highly detailed planet interior models, including
temperature structure and phase changes, are not necessary to derive
solid exoplanet bulk composition from mass and radius measurements.
For solid exoplanets with no substantial atmosphere we have also
found that: with 5\% fractional uncertainty in planet mass and radius
it is possible to distinguish among planets composed predominantly of
iron or silicates or water ice but not more detailed compositions;
with $\sim$~5\% uncertainty water ice planets with $\gtrsim 25\%$
water by mass may be identified; the minimum plausible planet size for
a given mass is that of a pure iron planet; and carbon planet
mass-radius relationships overlap with those of silicate and water
planets due to similar zero-pressure densities and equations of
state. We propose a definition of ``super Earths'' based on the
clear distinction in radii between planets with significant gas
envelopes and those without.

\end{abstract}


\section{Introduction}

The growing number and unexpected diversity of recently
discovered extrasolar planets has motivated us to study the
mass-radius relationship of solid exoplanets.  The
central question we pose is what can we determine about an
exoplanet's composition from its mass and radius? The answer to this
question requires numerical models of planet interiors as well as an
understanding of the current limitations and future prospects of the
precision of planet mass and radius observations.

The growing number of exoplanets includes some with interesting radii.
The planet HD~149026~b has such a small radius for its measured mass
that the planet must have a core with a mass of 60--70 $M_{\oplus}$,
or 2/3 of the planet's total mass \citep{sato2005}. Another planet,
GJ~436b, is a Neptune-mass planet \citep{butl2004} that was recently
discovered to show transits \citep{gill2007} and to have a
Neptune-like radius ($M_p = 22.6 \pm 1.9 M_{\oplus}$ and $R_p = 3.95
\pm 0.35 R_{\oplus}$).

The unexpected diversity of exoplanets includes fourteen exoplanets
with $M_p \sin i < 21 M_{\oplus}$ \citep{butl2004, mcar2004, sant2004,
rive2005, bonf2005, vogt2005, lovi2006, udry2006, bonf2007, melo2007},
including one with $M_p = 7.5 M_{\oplus}$ \citep{rive2005} and another
with $M_p = 5 M_{\oplus}$ orbiting at the inner edge of its host
star's habitable zone \citep{udry2007}. Microlensing surveys have
discovered two low-mass planets, $\sim 5.5 M_{\oplus}$
\citep{beau2006} and $\sim 13 M_{\oplus}$ \citep{goul2006} at
$\sim$~2.5~AU from their parent stars, suggesting that Neptune-mass
planets are common. In one planetary system, the Neptune-mass planet
is the most massive planet in the system \citep{goul2006}.

Space-based missions provide us with further motivation for our study.
COROT\footnote{http://smsc.cnes.fr/COROT/} (CNES; launched 27
December 2006) and Kepler\footnote{http://kepler.nasa.gov/} (NASA;
launch date 2008) will search for low-mass exoplanets that transit
their host
star. GAIA\footnote{http://sci.esa.int/science-e/www/area/index.cfm?fareaid=26}
(ESA; launch date 2011) will measure stellar distances (and hence
their radii) precisely, removing a limiting factor in deriving a
precise planetary radius. Ground-based radial velocity techniques are
pushing to higher precision, and will enable mass measurements of many
of the COROT and Kepler planets.

To examine what can we determine about an exoplanet's composition from
measurements of its mass and radius, we derive theoretical mass-radius
relationships for a wide range of exoplanet masses. To explore a wide
range of masses and compositions, we make a major simplification: we
make the approximation that the planet is at a uniform low
temperature. This approximation of uniform low temperature serves as a
practical simplification because the equations of state (EOSs) are
relatively well described at zero-temperature or at 300~K for a wide
variety of materials below 200 GPa. The full temperature-dependent
EOSs for the materials of interest are either unknown or highly
uncertain at the temperatures massive solid planets can reach in their
interiors and in the pressure range beyond the reach of static
compression experiments ($\lesssim$ 200 GPa) and the analytical high
pressure laws of plasma physics ($\gtrsim 10^4$~GPa).

We adopt our approach from the foundational work of \citet{zs1969}
who computed mass-radius relationships for homogeneous zero-temperature
spheres of single elements.  We improve upon \citet{zs1969} by using a
more accurate EOS at pressures $P \lesssim 1000$~GPa. We further
expand upon \citet{zs1969} by: considering more realistic planetary
materials; considering differentiated planets; by exploring the
effects of temperature on the planet mass and radius; and by
investigating potential observational uncertainties on planet mass and
radius.

This work complements the highly focused physical models of low-mass
exoplanets offered by other authors.  A detailed study by
\citet{lege2004} focused on water planets, and provided a detailed
model of the interior and atmosphere of a 6 $M_{\oplus}$ planet with
an interior composition of: 3 $M_{\oplus}$ of water, 2 $M_{\oplus}$ of
a silicate mantle and 1 $M_{\oplus}$ iron and nickel core (see also
\citet{kuch2003} for a description of a water planet.)
\citet{vale2006} calculated the mass-radius relationship for ``super
Earths'' and ``super Mercuries'' (defined in their paper to be 1--10
$M_{\oplus}$ and similar composition to Earth and 1--10 Mercury masses
and similar composition to Mercury, respectively).  \citet{vale2006}
explore different mineral compositions of the mantle and core, and
investigate whether the planets have solid or liquid cores and apply
their model to Gl~876d in \citet{vale2007a} and to degeneracies of
planet interior composition in \citet{vale2007b}.  \citet{ehre2006}
model small cold exoplanets to study the microlensing planet
OGLE~2005-BLG-390Lb \citep{beau2006}. In the past \citet{stev1982} and
more recently \citet{fort2007} and \citet{soti2007} have also
investigated mass-radius relationships of rocky and icy exoplanets.

We take a broader view than these previous studies using more
approximate models in order to investigate planets of a wide range of
compositions and masses.  We describe our model in \S2 and the
equations of state (EOS) in \S3.  In \S4 we present mass-radius
relationships for homogeneous and differentiated planets, and discuss
the effects of phase changes and temperature. In \S5 we describe a
generic mass-radius relationship shared by all solid exoplanets under our
approximation. We discuss the broad consequences of our study in \S6
followed by a summary and conclusion in \S7.

\section{Model}

We solve for $m(r)$, the mass contained within radius, $r$, $P(r)$,
the pressure, and $\rho(r)$, the density of a spherical planet
from three equations: the mass of a spherical shell
\begin{equation}
\label{eq:mass}
\frac{dm(r)}{dr} = 4 \pi r^2 \rho(r);
\end{equation}
the equation of hydrostatic equilibrium 
\begin{equation}
\label{eq:hydrostatic}
\frac{dP(r)}{dr} = \frac{-G m(r) \rho(r)}{r^2};
\end{equation}
and the equation of state (EOS)
\begin{equation}
\label{eq:eos}
P(r) = f(\rho(r),T(r)),
\end{equation}
where $f$ is a unique function for a given material. 
Different approximations to the EOS have been derived and we 
describe our choice in detail in \S\ref{sec-eos}. 
For the majority of the calculations for solid materials presented
here, we neglect the temperature dependence of the EOS and use
experimental data obtained at room temperature. The importance of
thermal contributions to the pressure are discussed in
\S\ref{sec-temperature}.

We numerically integrate equations (1) and (2) starting at the planet's center
($r=0$) using the inner boundary condition $M(0)=0$ and
$P(0)=P_{\rm central}$, where $P_{\rm central}$ is a chosen central
pressure.  For the outer boundary condition we use $P(R_p) = 0$.
The choice of $P_{\rm central}$ at the inner boundary and the outer
boundary condition $P(R_p)=0$ defines the planetary radius $R_p$ and
total mass $M_p = m(R_p)$.  Integrating equations (1) and (2) over and
over for a range of $P_{\rm central}$ provides the mass-radius
relationship for a given EOS.

For differentiated planets containing more than one kind of material,
we specify the desired fractional mass of the core and of each
shell. We then integrate equations~(1) and (2) as specified above,
given a $P_{\rm central}$ and outer boundary condition.  We switch from one
material to the next where the desired fractional mass is reached,
using a guess of the total planet mass.  Since we do not know the total
mass that a given integration will yield ahead of time, we generally need
to iterate a few times in order to produce a model with the desired
distribution of material.  

We tested our code by trying to duplicate the mass-radius curves in
\citet{zs1969}, using their EOS \citep{sz1967}.  Our mass-radius
curves agreed with those in \citet{zs1969} to within a few percent.

\section{Equations of State}
\label{sec-eos}
An equation of state (EOS) describes the relationship between density,
pressure, and temperature for a given material in thermodynamic
equilibrium.  Because we compute
models without temperature dependence, we choose a form of EOS that
assumes uniform or zero temperature.  For $P \lesssim 200$~GPa we use
fits to experimental data, either the Vinet EOS or the Birch-Murnagham
EOS (BME).  For $P \gtrsim$ 10$^4$~GPa, where electron degeneracy
pressure becomes increasingly important we use the Thomas-Fermi-Dirac
theoretical EOS. In between these pressures the EOSs are not well
known and we treat the EOS as described in \S\ref{sec-intermedEOS}.

\subsection{Low-Pressure EOSs: Vinet and Birch-Murnagham}

For pressures below approximately 200 GPa we rely on experimental data
which have been fit to the common EOS formulae of either Vinet
\citep{vin1987, vin1989} or BME \citep{birc1947, poir2000}. For a
derivation of these EOSs see \citet{poir2000}.  The Vinet EOS is
\begin{eqnarray}
P = 3K_0 \eta^{2/3}\left[1-
\eta^{-\frac{1}{3}}\right] 
\exp\left(\frac{3}{2}
(K'_0 - 1)\left[1- \eta^{-\frac{1}{3}}\right]
\right), 
\end{eqnarray}
and the third order finite strain BME is
\begin{eqnarray}
\label{eq:BME}
P = \frac{3}{2}K_0 \left[\eta^{\frac{7}{3}} - \eta^{\frac{5}{3}}\right]
 \left\{1 + \frac{3}{4} (K_0' - 4) \left[\eta^{\frac{2}{3}} - 1\right]\right\}. 
\end{eqnarray}
For the 4th order finite strain BME, the term
\begin{eqnarray}
+ \frac{3}{2}K_0 \left[\eta^{\frac{7}{3}} - \eta^{\frac{5}{3}}\right] 
\frac{3}{8} K_0 \left( \eta^{2/3} - 1 \right)^2 \\
\nonumber
\times 
\left[K_0 K_0'' + K_0'(K_0' - 7) + \frac{143}{9}\right]
\end{eqnarray}
is added to equation~(\ref{eq:BME}).
Here $\eta=\rho/\rho_0$ is the compression ratio with respect to the
ambient density, $\rho_0$.  $K_0=-V\left(\frac{\partial P}{\partial
V}\right)_T$ is the bulk modulus of the material, $K'_0$ is the
pressure derivative, and $K''_0$
is the second pressure derivative. The majority of experiments
(from which $K_0$ and $K'_0$ are derived) are typically limited to
pressures less than 150 GPa and temperatures less than 2000~K. 

Both the BME and the Vinet EOSs are empirical fits to experimental
data. The Vinet EOS is considered to be more suitable than the BME EOS
for extrapolation to high pressures, because the BME is derived from
an expansion of the elastic potential energy as a function of pressure
truncated to low orders \citep{poir2000}.  Where possible we choose
the Vinet or BME fit provided for experimental data according to which
fit best matches up with the TFD EOS at high pressures. In one case we
used a fourth order BME where the term $K_0''$ is determined
theoretically (see \S\ref{sec-intermedEOS}).  Table~1 lists the $K_0$,
$K_0'$, and the type of EOS fit we used for each material.

\subsection{High-Pressure EOS: Thomas-Fermi-Dirac}

The Thomas-Fermi-Dirac (TFD) theory was derived in the
late 1920s as an approximate way to characterize the interactions of
electrons and nuclei. The electrons are treated as a gas of
noninteracting particles that obey the Pauli exclusion principle and
move in the Coulomb field of the nuclei. Under the assumption that the
potential is slowly varying, a self-consistent solution is derived so
that Pauli exclusion pressure balances out the Coulomb
forces \citep{elie2002}.

These approximations lead to a comparatively simple description that
works for any material and becomes increasingly accurate at high
pressure.  For each material, however, there is a pressure limit below
which the TFD model is no longer valid, where the assumption of a
noninteracting electron gas in a slowly varying potential breaks
down. In real materials, the electrons occupy well-defined orbitals,
which leads to chemical bonds and determines the crystal structure as
a function of pressure and temperature. The TFD theory cannot describe
chemical bonds and is insensitive to the arrangements of atoms in a
particular structure. At very high pressure, however, where the kinetic
energy dominates over the Coulomb energy, all these effects become
less important and TFD theory yields an increasingly accurate EOS.

In this paper, we use a modified TFD theory developed by
\citet{sz1967}. The authors extended the original TFD theory by adding
a density dependent correlation energy term that captures some of the
interaction effects of the electrons. In all of the following TFD
calculations, we included the correlation energy correction calculated
with fit formulae provided in~\citet{sz1967}\footnote{Note
that the denominator in the second term of $\phi$ as defined in
\citet{sz1967} is listed as $4.3^{1/3}$; the factor $4 \times 3^{1/3}$
actually reproduces the EOSs in their paper. This definition of $\phi$
differs from the one in \citet{zs1969}, which appears to be missing a
factor of $1/3^{1/3}$.}. We also follow this paper for the
description of mixtures of different types of atoms.

\begin{figure}
\begin{center}
\includegraphics[width=14cm, height=14cm]{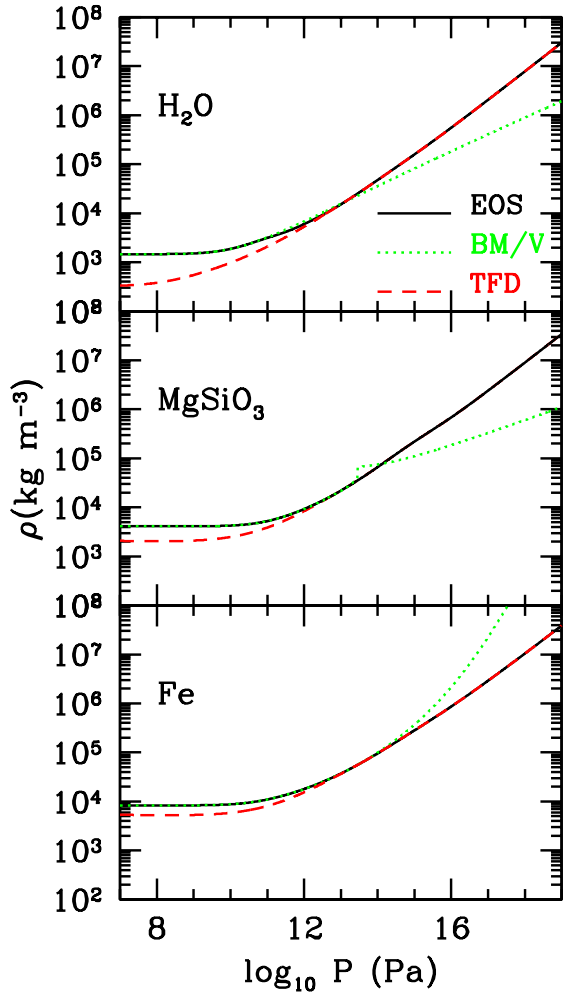}
\end{center}
\caption{Equations of state for different materials at zero or 300~K.
The black solid line is the EOS used in this study. The
green dotted line is a fit to experimental data (either Vinet (V) or
BME (BM)), appropriate for low pressures (typically below 200
GPa). The red dashed line is the TFD EOS, appropriate for high
pressures (typically above 10$^4$ GPa). We adopt the Vinet or BME EOS
at low pressures and switch to the TFD EOS at high pressures. Note
that the abrupt increase in density of the MgSiO$_3$ BME curve at $2.8
\times 10^{13}$ Pa is above the pressure where we switch over to the
TFD EOS, and illustrates the invalidity of extrapolating the BME to
high pressures. \label{fig:EOSmultia} }
\end{figure}

Since the TFD theory does not describe chemical bonds it does not
reproduce the correct zero-pressure density, and can even be in error
by up to a factor of two or more (\citet{zs1969}; see
Figure~\ref{fig:EOSmultia}). Instead we rely on
experimental data for lower pressure, which are avaliable for almost
all materials we consider.

\subsection{Intermediate-Pressure EOS and Details for Specific Materials Used}
\label{sec-intermedEOS}

The pressure range from approximately 200 to 10$^4$ GPa is not
easily accessible to experiment nor is it well described by the TFD
EOS. Although shock experiments can reach pressures over 1000 GPa, a
substantial contribution to the pressure comes from the thermal
pressure (see \S \ref{sec-temperature}) because the material is also
heated under shock compression. When a shock wave passes through a
solid material, its density is only increased by up to a factor of
$~4$. It is this limited compression ratio, which makes if very
difficult to obtain low temperature EOS data beyond 200 GPa, the range
needed to model planetary interiors.  For all materials but H$_2$O in
this pressure regime we simply use the Vinet/BME EOS up until the
point where it intersects the TFD EOS curve, and the TFD EOS at higher
pressures (see Figure~\ref{fig:EOSmultia}).
A more accurate EOS in the intermediate pressure range 200 to 10$^4$
GPa range demands new theoretical calculations. Such EOSs are not
readily available since there are almost no applications that require
EOSs in this pressure range.

{\bf Water Ice:} 
As an example of how to fill this gap we used density functional
theory to calculate the EOS of water ice in the phases VIII and X
in the pressure range 2 to 7700 GPa (Figure~\ref{fig:h2oEOS} and
Table~2). The theoretical EOS data presented here are in agreement
with water ice VII experimental data \citep{heml1987} in the range 6
to 127 GPa to within 3.5\% in density for a given pressure. Our Vinet
fit to the combined theoretical data for water ices VIII and X has
parameters $\rho_0 = 1460$~kg/m$^3$, $K_0$= 14.3771 GPa, and
$K'_0=6.57972$. This fit deviates from the tabular data in Table~2 by
less than 2.5\% in density for a given pressure.

At $P > $ 3 to 20 GPa water ice is in either phase VII or VIII,
depending on temperature (see the water phase diagram in, e.g., 
\citet{petr1999}). The structure of water ice phases VII and VIII are
extremely similar, differing only by the ordering of the hydrogen
atom. This different structure causes a negligible difference in the EOS,
making the experimental water ice VII and the theoretical VIII EOS
comparable. At $P=60$~GPa water ice VII/VIII undergoes a phase change
to water ice X. In this structure, the distinction between covalent
bonds and hydrogen bonds goes away. Instead the hydrogen atom shifts
to the mid-point between two oxygen atoms, while the hydrogen atoms
occupy off-center sites in all ice structures at lower pressures.
More details on our choices of EOSs follow.

For this work we adopt the following for our water
EOS. We use the BME fit from the \citet{heml1987} water ice VII data
up to $P=44.3$~GPa. At this pressure the theoretical data and the
experimental data agree precisely. Starting at $P=44.3$~GPa we use the
theoretical data derived from density functional theory, which
represent state-of-the-art first-principles calculations. These
calculations were performed with the Vienna {\it ab initio} simulation
package using the projector augmented-wave method
\citep{kres1996}. The calculations predict a gradual transformation
from ice VIII to X. The resulting EOS is given in Table 2. At a
pressure of 7686 GPa, our density functional theory calculation agrees
with the TFD model and we use this pressure to switch to TFD for all
higher pressures.  In principle, density functional theory
calculations can be performed for any material with a known crystal
structure---this provides a way to bridge the gap in pressure between
the experimental data and the TFD limit.

For the liquid water EOS for $P \leq 10$~GPa, we use the logarithmic
EOS \citep[see, e.g.,][]{poir2000}. We use $K_0 = 2.28$~GPa for seawater
at $12^{\circ}$~K from \citet{hali2003}.

{\bf Iron:} For an Fe EOS we use the $\epsilon$ phase of Fe with
a Vinet fit up to $P = 2.09 \times 10^4$~GPa.  Note that the Vinet fit
parameters are from experimental data with $P \leq 330$~GPa
\citet{ande2001}. At this pressure the Vinet curve smoothly approaches
the TFD EOS, and we switch to the TFD EOS.

{\bf Silicate:} For a silicate EOS we use the perovskite phase of
MgSiO$_3$. We use a fourth order BME fit up to $P = 1.35 \times 10^4$~
GPa.  At this pressure we switch to the TFD EOS.  The fourth order BME
fit is from a density functional calculation up to
$P = 150$~GPa by \citet{kark2000}. \citet{kark2000} note that their
$K'_{0}$ agrees with fits to experimental data from several sources
and $K_0$ is within the range of experimental data (247~GPa compared
to 246--272~GPa).  The advantage of the \citet{kark2000} fit
parameters is that the fourth order BME is the only fit we found that
smoothly matches the TFD EOS at high pressures.
 
{\bf Other materials:} Other materials used in this work include MgO
\citep{duff1995}, (Mg,Fe)SiO$_3$ \citep{knit1987}, and SiC
\citep{alex1989}.  For these species we use the BME fit up to the
pressure where they intersect the TFD curve.

Carbon was the only material we used whose Vinet EOS and TFD EOS did
not intersect; this was likely because we only considered the graphite
phase at pressures below the TFD EOS. For graphite we interpolated
between the Vinet and TFD EOS.  For H, He and the carbon monoxide EOS, we
used the TFD EOS at all pressures, for simplicity. We set the density
to a constant at the low pressures ($P < 10^8$ Pa) where the TFD EOS
is poorly behaved.  We note that a CO EOS from shock experiments
exists \citep{nell1981} in the pressure range 5 to 60 GPa, and it has
a density to within eight to sixteen percent of our TFD density. Our
CO EOS and mass-radius relationship is therefore approximate.

\section{Numerical Results}
\label{sec-results}

We now describe our numerical solutions to equations (1) through (3)
using our assembled collection of EOSs.  We used our model to
investigate the mass-radius relationships for planets from 0.01--1000
$M_{\oplus}$.  The lower mass limit encompasses planets as small as
Mercury and small bodies like the icy moons of Jupiter and Saturn.
The upper-mass limit encompasses the 13 Jupiter-mass planet
limit. Above this mass, self-gravitating H-He spheres undergo
deuterium or sustained hydrogen fusion (depending on how massive the
body is) and are not considered planets.

\subsection{Mass-Radius Relationships}

\subsubsection{Homogeneous Planets}
\label{sec-homoplanets}
Building upon \citet{zs1969}, we first consider planets of uniform
composition. This artificial scenario helps us understand the
fundamental properties of the planet mass-radius relationships.
Figure~\ref{fig:mrplanets} shows the mass-radius relationship for
homogeneous planets of H, H/He (25\% He by mass), H$_2$O (ice),
MgSiO$_3$ (perovskite), and Fe.

Homogeneous planets all show the same general trend in radius as a
function of mass. For $M_p \lesssim 500 M_{\oplus}$ the planets' radii
increase with increasing mass. In this regime, Coulomb forces balance
gravity in hydrostatic equilibrium.  For large masses, $M_p \gg 500
M_{\oplus}$, the compression in the interior is high enough to
pressure ionize the atoms. At these large masses degeneracy pressure
of free electrons balances gravity in hydrostatic equilibrium, and as
more mass is added to the planet, the planet shrinks \citep{hubb1984}.
Although planets are not fully degenerate (the term is reserved for
stellar mass white dwarfs \citep{chan1939}), electron degeneracy
pressure does have a significant effect on the mass-radius
relationship for high planetary masses over 500 $M_{\oplus}$. In
particular, planets of all compositions are approximately the same
size for a decade of mass where the competing effects of Coulomb
forces (which cause $R_p \sim M_p^{1/3}$) and electron degeneracy
pressure ($R_p \sim M_p^{-1/3}$) roughly cancel each other out.  See
\citet{zs1969} for a detailed discussion of the maximum radius for a
given planet of homogeneous composition.

If we assume that our selection of materials spans all plausible
major planet materials, then we can make some inferences from
Figure~\ref{fig:mrplanets} about the range of planet sizes.  First,
the Fe-planet mass-radius relationship shows the minimum radius a
planet of a given mass can posses.  Second, since water is the least
dense of all the materials we studied (apart from H and He), the water
planet curve in Figure~\ref{fig:mrplanets} may serve to show the
maximum radius for a planet with no substantial atmosphere.

The mass-radius relationships for planets of homogeneous compositions
(Figure~\ref{fig:mrplanets}) can be used to infer the bulk composition
of planets. Using the solar system as an example, and from
Figures~\ref{fig:mrplanets} and \ref{fig:mrplanetszoom}, we could
infer that Earth and Venus are composed primarily of a mixture of
silicates and iron, while Mercury is composed predominantly of
iron. We could also infer that Uranus and Neptune are not giant H/He
planets and nor are they ``rock giants''; they are predominantly rocky
or icy and must have small but significant gas envelopes. Jupiter and
Saturn are grossly fit by the H/He curve, but the H/He interiors of
hot Jupiters are dominated by thermal effects and are thus not fit
well by cold homogeneous planets; indeed we are not aiming to model
gas giant planets in this paper.

Turning to exoplanets, Figure~\ref{fig:mrplanets} shows that the
transiting Saturn-mass\footnote{Saturn is 95~$M_{\oplus}$ and Jupiter
is 318~$M_{\oplus}$.} exoplanet HD~149026b must contain a substantial
fraction of elements heavier than H and He.  More detailed
evolutionary and interior models find that HD~149026b has 70
$M_{\oplus}$ of rocky material, almost 2/3 of its total mass
\citep{sato2005}. At 22.6 $M_{\oplus}$ and 3.95 $R_{\oplus}$, GJ~436b
must have a significant H/He envelope because its radius is
clearly larger than a pure water ice planet. The 7.5 $M_{\oplus}$ planet
Gliese 876d \citep{rive2005} does not have a measured radius, but a
radius measurement with a fractional radius uncertainty of 5\% would
distinguish among a predominantly rocky planet, a predominantly icy
planet and a planet with a substantial gas envelope.

We have modeled massive solid exoplanets up to 4000 $M_{\oplus}$ (up
to 13 Jupiter masses). These planets would be Jupiter-mass planets
composed of solid material. Such massive exoplanets are not yet known
to exist. In the standard planet formation theory massive planets are
primarily composed of H/He and are limited to have rocky/icy cores of
up to about $10 M_{\oplus}$. HD~149026b with a 70~$M_{\oplus}$ core
shows that a wider range of planets exist. Massive solid exoplanets of
hundreds to thousands of Earth masses may be able to form around
massive stars (B and O stars; 5--120 $M_{\odot}$) where the
protoplanetary disk would contain enough heavy elements. Additionally
these stars have high UV radiation and winds which could
photoevaporate the nascent protoplanetary gas disk, allowing massive
planets to form out of the remaining solid material.

\subsubsection{Differentiated Planets}
\label{sec-diffplanets}

All solar-system planets have multiple layers of different
compositions. These planets are differentiated, meaning
the denser material lies beneath shells of progressively less
dense material.  We now consider differentiated planets of
various compositions.  We focus on materials that comprise the solar
system planets and moons: iron, silicates, H$_2$O (ice), and H/He gas
envelopes. We ignore elements that have abundances too low to affect
our model for the planet radius.

We explore two types of differentiated gas-free planets, iron/silicate
planets and water planets.  These planets lack gas envelopes, although
they may have atmospheres too small to affect the measured planet
radius.  Figure~\ref{fig:mrplanets} shows the mass-radius relationship
for differentiated planets without gas envelopes.  Figure
\ref{fig:mrplanetszoom} shows the same mass-radius relationships in
more detail.  The calculations assume a constant fractional mass in
each layer.  In general, the radii of differentiated planets (where
the more dense components are interior to the less dense components)
lie in between the radii of homogeneous spheres composed of the
planet's most and least dense components.

We investigate iron/silicate planets with iron cores and
MgSiO$_3$ mantles.  We consider Fe core mass fractions of 32.5\%
(``super Earths'') and 70\% (``super Mercuries''). It is remarkable
how well Mercury, Venus, and Earth's masses and radii are fit by these
cold iron/silicate planets, as seen in Figure~\ref{fig:mrplanetszoom}.
We show the density as a function of radius for the silicate planets
in Figure \ref{fig:rhocurves}a and b. As expected, the more massive
planets have higher densities in their cores. 

We define water planets to be solid planets with $> 25$\% H$_2$O by
mass. Jupiter's moons Ganymede and Callisto would be water planets by
this definition\footnote{Under a water planet definition that includes
H/He envelopes, Uranus and Neptune would also be water planets,
because they are believed to have $> 25$\% water by mass.}.  We
investigate water planets with iron cores and silicate mantles. We
consider planets with fixed mass fractions: 45\% water, 48.5\%
silicates and 6.5\% Fe (similar to Jupiter's icy moon Ganymede
\citep{schu2004}); 75\% water and 22\% silicates and 3\% iron (the
core and silicate shell mass ratio as Ganymede); and 25\% water with a
58\% silicate shell and a 17\% iron core.  Figure~\ref{fig:rhocurves}c
shows the density profiles as a function of planet radius for the 45\%
water planets.

We compared our fiducial super-Earth model with a 32.5\% Fe core and
67.5\% silicate mantle to a model that more closely represents Earth:
a model with a 32.5\% by mass core of FeS (where FeS includes 10\% S
by mass) and a mantle that includes 90\% (Mg,Fe)SiO$_3$ and 10\% MgO
\citep{poir2000}. The results from this Earth model are shown by the
squares in Figure~\ref{fig:phase} and agree closely with our fiducial
super-Earth-like planet.  It is remarkable how our simple fiducial
model matches the Earth's radius to within three percent for a $1
M_{\oplus}$ planet.

There are further degeneracies among the mass-radius relationships for
planets of different compositions. For example, planets with 10\%
water by mass (with 27\% iron cores and 63\% silicate mantles) have
mass-radius curves that overlap with our silicate planet mass-radius
curves. As a second example, if we adopt a 6 $M_{\oplus}$ water planet
similar to the \citet{lege2004} water planet composed of 1
$M_{\oplus}$ Fe, 2 $M_{\oplus}$ silicates, and 3 $M_{\oplus}$ H$_2$O,
we find the total planet radius differs by less than 0.5\%
from that of our model with 45\% water, 48.5\% silicates, and 6.5\%
Fe.  While the total planet mass and radius are the same, the interior
structure of the models are quite different, as shown in
Figure~\ref{fig:phase}d (but note that the interior structure of our
Leger-type planet is different than the one in \citep{lege2004}). For
the same mass, the radius of the Leger-type planet will be slightly
lower, but both types of planets fall along the same mass-radius
curve.  See \citet{vale2007b} and Li et al. (in prep) for detailed
discussions on degeneracies in planets composed of iron cores,
silicate mantles, and water outer layers.

\subsubsection{Planets with H/He Gas Envelopes}

We now turn to a discussion of differentiated planets with significant
H/He envelopes.  For simplicity and consistency, we use a
zero-temperature EOS for H and He \citep{sz1967}.  A
zero-temperature EOS for a H/He mixture may represent real planets
only poorly, but since a zero temperature EOS underestimates a gas's
volume, using such an EOS allows us to one important point: adding a
H/He shell can easily boost a planet's radius dramatically.  For
example, adding 20\% H/He by mass can double a planet's radius.

Figure~\ref{fig:gasplanets} shows mass-radius relationships for
planets with gas envelopes and fixed core masses.  The cores contain
70\% MgSiO$_3$ and 30\% Fe. Five cases are shown: planets with fixed
core masses of 5, 10, 20, 50, and 100 $M_{\oplus}$. The values in
Figure~\ref{fig:gasplanets} are lower limits because thermal effects
will move the gas curves up and left, i.e. to larger radii.  For a
detailed discussion of mass-radius curves for planets with significant
gas envelopes see \citet{fort2007} and Adams et al. in preparation.
In Figure~\ref{fig:gasplanets} we also show the radius of the core of the
H/He planets as a function of the total mass of the fixed-mass core
and overlying H/He envelope.

All planets with a significant amount of H/He will have a larger
radius than the homogeneous water ice planets. Planets with even a
relatively small fractional mass of H/He can therefore be
distinguished from planets with an insignificant gas envelope
(Figure~\ref{fig:gasplanets}). Uranus and Neptune, for example, are
believed to have about 10\% of their mass in H/He material. Yet,
Uranus and Neptune are up to two times larger than a planet of similar
mass and core composition but without the H/He envelope.  
We propose
to call planets without a significant gas envelope ---i.e. planets that lie
below the pure water ice line---super Earths.

\subsubsection{Non-Standard Planets: Carbon and Helium Planets}

We now consider planets with compositions very different from solar
system planets, beginning with carbon planets.  We have previously
presented the idea of carbon planets, planets composed of $>$ 50\%
carbon compounds by mass (\citet{kuch2005}; See also \citet{came1988}
and \citet{gaid2000}).  Carbon planets should form in environments
where the carbon-to-oxygen ratio C/O $\gtrsim$ 1, in contrast to the
solar abundance ratio (C/O = 0.5).  When C/O $\gtrsim$ 1, the
high-temperature condensates available in chemical equilibrium are
very different from the environment with C/O $<$ 1 \citep{lewi1974,
wood1993, lodd1997}.  For example, SiC is the dominant form of Si
instead of silicates (i.e., Si-O compounds).  Such carbon-rich
environments may occur in a local area enriched in C or depleted in
H${}_2$O in an otherwise solar-abundance protoplanetary
disk. Carbon-rich environments would also occur in a protoplanetary
disk with a global C/O $>$ 1 like the disk that formed the planets
around pulsar PSR 1257+12 \citep{wols1992}, and do occur in the
well-known Beta Pictoris debris disk \citep{robe2006}.

The dominant composition of carbon planets is unknown.  The details
would depend on protoplanetary disk temperature, composition, relative
abundance and departures from chemical equilibrium. SiC and graphite
are both plausible compositions (based on observations and
calculations of carbon star atmospheres; \citet{lodd1997}). We compute
the mass-radius relationship for three different kinds of carbon
planets.  The first carbon planet is a planet with an Fe core and an
SiC mantle. The second is a carbon planet with an Fe core and a
graphite mantle. The third type of carbon planet is a pure CO (carbon
monoxide) planet. A carbon monoxide planet could form in a stellar
disk composed from a CO white dwarf that has been shredded by its more
massive stellar binary companion \citep{livi1992}.  

The main result from our carbon planet computations
(Figure~\ref{fig:carbonplanets}) is that the carbon planet mass-radius
relationships overlap those of the silicate and water planets. This is
because the zero-pressure density of SiC (3.22 g/cm$^3$) is similar to
that of MgSiO$_3$ (4.10 g/cm$^3$). While the zero-pressure density of
graphite (2.25 g/cm$^3$) is almost twice that of water ice VII (1.46
g/cm$^3$), a graphite planet with an Fe core has a similar average
density to a water planet with an iron core and silicate mantle. CO
planets' mass-radius curves also overlap the mass-radius curves of
water planets that contain iron and silicate because of a similar
zero-pressure density. Our CO planets are similar in average density
to water ice planets, again because the zero pressure
densities are similar. The precise mass-radius relationship of CO planets
should involve temperature, and a more accurate EOS than
our adopted one.

We also compute the mass-radius relationship for a pure He planet.  We
again use the zero-temperature He EOS \citep{sz1967} to avoid
introducing a free parameter based on the planet's unknown entropy.
The helium mass-radius relationship is therefore approximate. 

A predominantly He planet may potentially form from a low-mass white
dwarf. For example, a He planet can conceivably form in one type of
symbiotic binary star called an AM CVn (AM Canes Venatici), composed
of two very H-poor white dwarfs (i.e., He core white dwarfs)
surrounded by a circumbinary helium accretion disk formed during mass
transfer from the less massive to the more massive white dwarf (see,
e.g., \citet{pods2003}, and references therein).  After it loses most
of its mass, the less-massive WD may approach planetary mass.

\subsection{Phase Changes and Thermal Effects}
\label{sec-phasetemp}
Real planets have phase changes and temperatures above 300~K in their
interiors.  Here we investigate the effect of phase changes and
temperatures on the planet mass and radius.  While models
that include temperature and phase changes can be more realistic than
our simple models, they complicate the interior boundary conditions
and necessarily involve regimes where the EOSs are highly uncertain.
See \citet{lege2004, vale2006, sels2007, soti2007, vale2007a} for low-mass
planet interior models that do include phase and temperature changes
for specific types of planets.

\subsubsection{Phase Changes}
\label{sec-phase}

In this section we show that the low-pressure phase changes have
little effect on the planet mass-radius curves.  This lack of effect
is because the low-pressure phase changes occur at pressures $<$ 10
GPa.  Figure~\ref{fig:fracmassradius} shows the pressure at $m(r) =
0.97 M_p $ for homogeneous planets.
Figure~\ref{fig:fracmassradius} shows that for planets $ \gtrsim 3
M_{\oplus}$, most of the planet's mass is at pressures higher than 3
GPa (with the exception of pure water planets).  Here we discuss the
phase changes in some of the key planetary materials individually.

{\bf Water:} We begin by investigating the liquid water to water ice
phase change. The room-temperature bulk modulus $K_0$ for liquid water
is an order of magnitude smaller than $K_0$ for water ice
VII. Furthermore, the densities of liquid water and water ice VII are
different by 50\% (see Table~1).  Despite these large differences in
physical properties, it is important to remember that liquids are
still not highly compressible materials and we expect them to behave
more similarly to solids than to gases under high pressure.  In order
to investigate the effect of a liquid water phase for a water planet,
we consider liquid water to be present at $P < 10$~GPa. The
temperature would be $\sim$~650~K for water to remain a liquid at
this high pressure.  We consider the differences between a pure water
ice planet with and without the liquid ocean. We find less than three
percent difference in the radius for water ice planets without and
with a liquid water ocean for planets in the mass range 1 to 4
$M_{\oplus}$.  We find $<1$\% difference in the radii of the two types
of planets for $M_p$ above 5 $M_{\oplus}$ (see
Figure~\ref{fig:phase}).

{\bf Silicates:} We now turn to phase changes in our MgSiO$_3$
perovskite planets by considering the low-pressure phase of MgSiO$_3$
called enstatite.  We adopt the low-pressure enstatite phase at
pressures less than 10 GPa. The difference in $K_0$ and $\rho_0$
between the high-pressure perovskite phase and the low-pressure
enstatite phase are much less than the differences between liquid water
and water ice (see Table~1). Hence, just as in the water planet case,
we see a relatively small difference ($< 1$\%) in radius
between silicate perovskite planets with and without a layer of
enstatite at $P \leq 10$~GPa (see Figure~\ref{fig:phase}).

{\bf Iron:} We now turn to phase changes in Fe.  The low-pressure
phase of Fe is Fe($\alpha$). Fe($\varepsilon$) and Fe($\alpha$) have
the same $K_0$ and $K_0'$ to within the experimental error bars, and
only slightly different zero-pressure densities (see
Table~1). Additionally, Fe ($\alpha$) only exists at $<$ 10 GPa;
whereas in a differentiated planet Fe is expected to exist at
pressures higher than 10 GPa and in a homogeneous Fe planet most of
the planet's mass is above 10 GPa (Figures~\ref{fig:mrfn} and
\ref{fig:fracmassradius}).  For these reasons the phase change of
Fe($\alpha$) to Fe($\varepsilon$) has little to no effect on the
mass-radius relationship of Fe planets.

{\bf High pressure phase changes:} Even if there is a phase change at
high pressures (not considered here), we expect the associated
correction to the EOS to be small, and hence the derived planet radii
to be reasonably accurate. Phase changes arise from a rearrangement of
atoms in the crystal structure and the associated modifications in
chemical bonds. With increasing pressure, the atoms become packed more
and more efficiently and the importance of chemical bonding patterns
drops significantly. For example, above 100 GPa, the postperovskite phase
\citep{mura2004} and perovskite phase of silicate have very similar
zero pressure volumes and bulk moduli, to less than a few percent
difference \citep{tsuc2005}. We caution that although phase changes
do not need to be considered, theoretical calculations are needed to
compute an accurate EOS at high pressures (i.e., between 200 and
10$^4$ GPa) because in many cases an extrapolation of the BME or Vinet
EOS fit will not do (see \S\ref{sec-intermedEOS}). For example, for
some water ice planets with masses above 10 $M_{\oplus}$, the effect
of extrapolating the BME fit without considering a more accurate
higher pressure EOS is considerable (Figure~\ref{fig:phase}d).

At very high pressures the TFD EOS becomes valid. To understand the
planet mass range where TFD effects are important, we show planet
mass-radius relationships in the case of ignoring the TFD EOS in the
dotted lines in Figure~\ref{fig:phase}, i.e. in using only the
extrapolated BME or Vinet fit.  

\subsubsection{Temperature}
\label{sec-temperature}

Temperature has little effect on the radius of solid exoplanets.
The reason is that the density of a solid changes by a relatively
small amount under the influence of thermal pressure. At low
pressures, the crystal lattice structure dominates the material's
density and the thermal vibration contribution to the density is small
in comparison.  At high pressures, the close-packed nature of the
atoms prevents any significant structural changes from thermal
pressure contributions. Moreover, any change in average density of a
planet of fixed composition results in a smaller change in the planet
radius than the change in average density, because $R_p \sim
\bar{\rho_p}^{-1/3}$.

Some authors have stated or shown with models that temperature is not
important for deriving a planet's total mass and radius \citep[see
e.g.,][and references therein]{vale2006, vale2007a, fort2007,
soti2007}.  In this subsection we go into detail to estimate the
effect of temperature on density for the three main materials that we
studied, iron, silicates, and water ice VII.

Heating planetary material adds a thermal contribution to
the pressure term. In a first approximation
one can assume that the thermal pressure is a linear
function of temperature
\citep[e.g.,][]{ande1989, poir2000}.  The thermal
contribution to the pressure, $P_{th}$, can be written as:
\begin{equation}
P_{th} = \int_0^T \left(\frac{\partial P}{\partial T} \right)_V dT = \int_0^T \alpha(T) K_0(T) dT. 
\end{equation}
Here $\alpha(T)$ is the thermal expansivity and $K_0(T)$ is the bulk
modulus as before. 

For most cases the term $\alpha(T) K_0(T)$ is independent
of volume above the Debye temperature $\Theta_D$ \citep{ande1989}.
The thermal pressure term can then be written
\begin{equation}
P_{th} = \int_0^{\Theta_D} \alpha K_0 dT + \alpha K_0 (T - \Theta_D).
\end{equation}
Note that the $\Theta_D$ is around 700K for silicate- and Fe-bearing
minerals.  $\Theta_D$ is significantly lower (300K) for water ice and
significantly higher for carbon (2000K).  Moreover, for Earth-bearing
minerals \citet{ande1989} find that $P_{th}$ is linear in $T$ down to
much lower temperatures, and find that
\begin{equation}
\label{eq:Pth}
P_{th} = \alpha K_0 (T - 300~{\rm K}).
\end{equation}
For (Mg,Fe)SiO$_3$ we use $\alpha K_0 = 0.00692$~GPa/K from
\citet{ande1994}. We note that, for our purposes, (Mg,Fe)SiO$_3$ is
similar enough to MgSiO$_3$ for estimating the thermal pressure.

For metals the thermal excitation of electrons must be taken into
account by including a higher order term \citep[][and references
therein]{isaa2003},
\begin{equation}
\label{eq:Pth2}
P_{th} = \alpha K_0 (T - 300~{\rm K}) + \frac{\partial \alpha
K_0}{\partial T}_V (T - 300~{\rm K})^2.
\end{equation}
We use values of $\alpha K_0 = 0.00121$~GPa K$^{-1}$ and
$(\partial \alpha K_0) / (\partial T)_V = 7.8 \times
10^{-7}$~GPa K$^{-2}$ from \citet{isaa2003}.

For H$_2$O ice VII, we constructed $P_{th}$ based on (P,V) isotherms
according to the thermodynamic relations in equations~(2) and (3) in
\citet{fei1993}.  This method uses a linear fit with $T$ to
$\alpha(T)$ and a parameter $\eta_1$ which describes the pressure
effect on the measured volume at high temperature. We take the
parameters from \citet{fran2004}.

Figure~\ref{fig:Pth} shows $P_{th}$ vs. temperature for Fe, H$_2$O ice
VII, and (Mg,Fe)SiO$_3$.  $P_{th}$ increases more slowly
at high pressure than at low pressure because at high pressure
the thermal expansion becomes a constant; the atoms become
tightly packed so that any thermal pressure has a decreasing contribution to
the total pressure. Although water ice data is only available for ice
VII to 50 GPa and $\sim$~800 K \citep{fran2004}, we expect it to show
the same $P_{th}$ trend with increasing temperature as other
materials.

We can estimate the change in density caused by a thermal pressure. We
computed $\rho(T)$ corresponding to the total pressure $P_{total} = P
+ P_{th}$ using the EOSs described in
\ref{sec-intermedEOS}. Figures~\ref{fig:Pth} to \ref{fig:Pthmg} show
($P$,$\rho$) isotherms for the main materials we studied.

For MgSiO$_3$, the decrease in density due to thermal pressure is less
than 4 percent over the temperature range 300~K to 6000~K and above 10 GPa
where the material is solid (Figure~\ref{fig:Pthmg}a). Below 10 GPa
the decrease in density is less than 2.5 percent for temperatures up
to 1200~K (Figure~\ref{fig:Pthmg}b and for densities less than the
zero-pressure 300~K density see Figure~7 in \citet{ande1994}.)

Fe is known to have a higher thermal expansivity than MgSiO$_3$ and
its density therefore decreases more than MgSiO$_3$ for the same
temperature increase.  The Fe density decrease is less than four
percent for pressures above 100 GPa
(Figure~\ref{fig:Pthfeandh2o}a). Earth's Fe core is at pressures higher
than 100 GPa \citep{prem1981}, and we expect the same for more massive
differentiated planets with similar composition (see,
e.g. \cite{vale2006}). We further note that as the planet's Fe mass
fraction increases, more of the planet's mass is at higher pressures,
making the material's density change at low pressure in response to
temperature less relevant. For example,
Figure~\ref{fig:fracmassradius} shows that for a pure Fe planet with
$M_p > 0.1 M_{\oplus}$, 97 percent of the planet's mass is above 1 GPa
and for planets with $M_p > 1 M_{\oplus}$ 97 percent of
the planet's mass is above 10 GPa.

Using available data to 50 GPa and 800 K \citep{fran2004}, we find
that H$_2$O water ice VII density changes by less than a few percent
throughout this pressure and temperature range
(Figure~\ref{fig:Pthfeandh2o}b). We note that above approximately
1000~K water ice reaches an ionic fluid phase \citep{gonc2005}.
Nevertheless, we emphasize that our water ice EOS (which includes
phases VII, VIII, and X; see Table~2) agrees with recent Hugoniot
shock data to within the experimental uncertainties. This recent data
is from \citet{lee2006} and ranges in pressure from 47 to 250 GPa and
temperature from 2100~K all the way to 19,000~K.  More work needs to
be done to quantify the thermal pressure effects above 250~GPa and in
the ionic phase, which is beyond the scope of this paper.

Our ($P$, $\rho$) isotherms in Figures~\ref{fig:Pthmg} and
\ref{fig:Pthfeandh2o} can be interpreted in light of the
temperature-pressure profiles of planetary interiors.  Earth, for
example, has temperatures of approximately 1000~K at 50 GPa, 2000~K at
150 GPa and up to 6000~K at 350 GPa \citep{poir2000}. Super-Earth
interior models calculated in \citet{vale2006} shows that below 1~GPa
the temperature is less than 1600~K, above 100 GPa the temperature is
2500~K, and the temperature is up to about 7000~K at hundreds of GPa.
Isentropes for Neptune are expected to be about 2500~K at 20 GPa and
to reach 7000~K at 600~GPa. The fact that high temperatures are
reached only at high pressures lessens the temperature effect because
the fractional contribution of the thermal pressure to the total
pressure decreases as the total pressure increases
(Figure~\ref{fig:Pth}).

We argue that even for the short-period hot exoplanets---those in few
day orbits whose surface temperatures could reach 1000 to 2000~K---the
thermal pressure contribution to radius is still small. This statement
is based on the argument that for solid planets above one to a few
Earth masses (depending on composition) the fractional radius affected
by such high temperatures is small.

Based on the above discussion of density decrease as a function of
temperature we illustrate the effect of temperature on planet radius
by the following example.  Taking a case where the average density is
overestimated by 3.5 percent in a uniform 300~K temperature models,
the planet's total radius would be underestimated by only 1.2
percent. This is due to the $R_p \sim \bar{\rho_p}^{-1/3}$ scaling.  

\section{A Generic Mass-Radius Relationship for Solid Exoplanets}
\label{sec-genericmr}

A glance at Figure~\ref{fig:mrplanets} may suggest to the astute
reader that the mass-radius relationships for a variety of planets all
have a similar functional form, perhaps caused by some symmetry of the
underlying equations.  Indeed, there is such a symmetry, and a common
functional form: a generic mass-radius relationship which we describe
here.  This generic mass-radius relationship is valid for planet
masses up to about 20~$M_{\oplus}$.

\subsection{A Modified Polytropic Equation of State}
\label{sec-modifiedpolytrope}
Our generic mass-radius relationship is based on the similar forms of the
EOSs of all solid materials we have considered.
The zero-temperature or 300~K temperature EOSs for
the solid materials we considered can be
approximated by the function 
\begin{equation}
\label{eq:modifiedpoly}
\rho(P) = \rho_0 + cP^n.  
\end{equation}
Table~4 lists the
best-fit parameters $\rho_0$, $c$, and $n$ for some materials over
the range $P < 10^{16}$ Pa.  The density given by these approximate
EOSs match the density given by the more detailed EOSs we used above to
within 2 to 5\% for the pressure ranges $ P < 5 \times 10^{9}$ and $P > 3
\times 10^{13}$ Pa.  At intermediate pressures, the discrepancy ranged
from less than 1\% to 12\%. 

The similarity of the EOSs of all solid materials stems from the
the behavior of chemical bonds under pressure. At low pressures the
chemical bonds of the material can withstand compression.  Above some
pressure, the energy imparted to the material breaks the chemical
bonds and the material structure radically changes. The ``crossover''
pressure is roughly the material's bulk modulus.  For our simple
analytic EOS, the bulk modulus is
\begin{equation}
K \approx \left( \frac{\rho_0}{c} \right)^{\frac{1}{n}}.
\end{equation}

Our approximate EOS is a modified polytropic EOS. Polytropic EOSs are
of the form $P = K_p \rho^{1+1/m}$, or $\rho = (P/K_p)^{m/(m+1)}$,
where $K_p$ is a constant and $m$ is the polytropic index.  Our
approximate EOS fit differs from the polytropic EOS by the addition of
the constant $\rho_0$; it incorporates the approximate incompressibility
of solids and liquids at low pressures.

\subsection{Dimensionless Equations of Planetary Structure}

The new generic EOS contains some dimensional quantities, $\rho_0$ and
$c$, that allow us to conveniently write equations~(1) and (2) in
dimensionless form.  We rescale the variables $P$, $\rho$, $m$, and
$r$ as follows, where the subscript $s$ refers to a scaled variable,
\begin{equation}
\label{eq:scalerho}
\rho_s = \frac{\rho}{\rho_0}
\end{equation}
\begin{equation}
\label{eq:scaledP}
P_s =  \frac{P}{P_1} 
\end{equation}
\begin{equation}
\label{eq:scaledR}
r_s = \frac{r}{r_1} \\
\end{equation}
\begin{equation}
\label{eq:scaledM}
\label{eq:scalem}
m_s = \frac{m}{m_1} \\
\end{equation}
where
\begin{equation}
P_1 =  \left(\frac{\rho_0}{c}\right)^{1/n}  \\
\end{equation}
\begin{equation}
r_1 = G^{-1/2} \rho_0^{(1/2n-1)} c^{-1/2n} \\
\end{equation}
and
\begin{equation}
m_1 = {r_1^3}{\rho_0}. \\
\end{equation}
With this change of variables, the equations of planetary structure become:
mass of a spherical shell
\begin{equation}
\label{eq:scaledmasseq}
\frac{d m_s(r)}{d r_s} = 4 \pi r_s^2 \rho_s(r_s);
\end{equation}
hydrostatic equilibrium
\begin{equation}
\label{eq:scaledhseeq}
\frac{d P_s (r_s)}{d r_s} = -\frac{\rho_s(r_s) m_s(r_s)}{r_s^2};
\end{equation}
and the EOS
\begin{equation}
\label{eq:scaledrhoeq}
\rho_s = 1 + P_s^n.
\end{equation}

Figure~\ref{fig:mrscaled} shows dimensionless mass-radius
relationships derived by numerically solving the above equations the
same way we solved the unscaled equations.  It shows the total
dimensionless planet mass, $M_s$ as a function of the scaled planet
radius, $R_s$, where $R_s$ is defined by the outer boundary condition
$P(R_s) = 0$. At $R_s$ we also have $M_s = m_s(R_s)$.  We generated
the numerical mass-radius curves by solving the equations for a range
of central pressure values of $P_s$.

The dimensionless mass-radius curves depend only on $n$.  We plot
curves for three values of $n$: $n=0.513$ (H$_2$O), $n=0.528$ (Fe),
and $n=0.544$ (silicate).  These values of $n$ span the range of
behaviors of all the EOSs we studied. We also show solutions for the
interior structure of homogeneous planets in
Figure~\ref{fig:rhoscaled}.  

The solutions behave quite differently on either side of the line $M_s
=1$.  For $M_s < 1$, $R_s$ strictly increases with $M_s$, and does not
depend on $n$.  For $M_s > 1$, $R_s$ depends strongly on $n$ and does
not necessarily strictly increase with $M_s$.  This contrasting
behavior arises because for $M_s \ll 1$, the EOS reduces to $\rho_s = 1$,
and for for $M_s \gg 1$, the EOS reduces to the polytropic form
$\rho_s = P_s^n$.

Table~4 lists some values of $m_1$, $r_1$, and $P_1$ for some EOSs
based on the values of $\rho_0$, $n$, and $c$ listed in Table~3.
Curiously, the scaling parameters $m_1$ and $r_1$ are somewhat similar for
the polytropic-like mass-radius solutions for H$_2$O and Fe because
these materials have similar ratios of $K_0 / \rho_0 $.

For $M_s < 4$, the dimensionless mass-radius relationship is
approximately
\begin{equation}
\log_{10} R_s = k_1 + 1/3 \log_{10}(M_s) - k_2 M_s^{k_3},
\label{eq:scaledmrapprox}
\end{equation}
where $k_1 = -0.20945$, $k_2 = 0.0804$, and $k_3 = 0.394$; this
approximation is good to 1\% over this range.  For $M_s > 4$, the
scaled radius becomes a strong function of $n$.  But We can use the
analytic function in equation~(\ref{eq:scaledmrapprox}) to describe the
dimensionless mass-radius relationships for scaled masses, $M_s$, up
to $\sim 40$ if we use the values of $k_i$ for
different materials listed in Table~4.

Equation~(\ref{eq:scaledmrapprox}) and the scaling parameters in
Table~4 provide a convenient approximate summary of the results of
this paper for homogeneous planets.

It may appear that the scaled mass-radius relationship is not useful
for differentiated planets, since differentiated planets combine
materials with different equations of state.  We find, however, that
even the radii of differentiated planets are well described by the
functional form shown in Figure~\ref{fig:mrscaled} for $M_s \lesssim
4$ (i.e., up to where the scaled numerical solutions for different
materials differ from each other).

This circumstance provides us with a convenient way to summarize our
results for differentiated planets.  To any differentiated planet
model, we can assign an effective $m_1$ and $r_1$.  These effective
scalings allow us to summarize {\it all} of our calculated mass-radius
relationships using equation~(\ref{eq:scaledmrapprox})---regardless of
whether they were computed with the modified-polytropic EOS or our
actual Vinet/BME and TFD EOSs. Table~4 lists some effective values of $m_1$ and $r_1$
for a few examples of differentiated planets.  We calculated these
numbers by comparing the mass-radius curves for differentiated planets
(shown in Figure~\ref{fig:mrplanets}) to
equation~(\ref{eq:scaledmrapprox}). To find the approximate
mass-radius relationship for any given planet, look up $m_1$ and $r_1$
in Table~4 and plug these numbers into
equation~(\ref{eq:scaledmrapprox}) using $k_1 = -0.20945$, $k_2 =
0.0804$, and $k_3 = 0.394$.

The scaled variables help us to understand why the solid planet
mass-radius relationships are very nearly the same for all materials
we considered, for $M_s <= 1$.  We first recall that $ M_s = 1$ is
defined as the mass so that the central, i.e. the maximum, pressure in
the planet is $P_s=1$. Next, for $P_s <= 1$, i.e. everywhere in the
planet for $M_s <=1$, the density (or the scaled density) never
changes by more than 2.5 percent if a variation of $n$ in the range of 0.513
to 0.549. Hence the radius (or the scaled radius) never changes by
more than 0.85 percent (due to the $R_p \sim \rho ^{-1/3}$ scaling).

\subsection{Analytic Treatment of the Dimensionless Equations}
Here we will derive an approximate analytic solution to the
dimensionless equations of planetary structure. The existence and form
of the solution demonstrates why the mass-radius curves for various
planets all look so similar.  The good agreement between this
approximate solution and our calculations gives us confidence in our
results.

We first discuss a general analytic solution to equations (1) and (2),
followed by an application to the dimensionless equations
(\ref{eq:scaledmasseq}) and (\ref{eq:scaledhseeq}). For most equations
of state, equations (1) and (2) cannot be solved analytically, even
given the approximation of zero temperature.  We can incorporate two
ideas to allow new analytic progress.  First, over a wide range of low
pressures below a GPa, solids and liquids change their densities by a
small amount---much less than ten percent. This point enables us to
assume an equation of state that incapsulates the idea of materials
that are largely incompressible over a wide range of low pressures:
\begin{equation}
\rho (P) =\rho_0 + f(P).
\label{eq:stateepsilon}
\end{equation}
We have shown in \S \ref{sec-modifiedpolytrope} that $f(P) = cP^n$ is
a good approximation.  We further note, however, that in the range of
low pressures we can assume $f(P) \ll \rho_0$.

The second point that enables an analytic treatment is that, when a
planet is massive enough that it begins to compress under its own
gravity, the compression is most acute at the planet's center.  With
these two ideas in mind, we can obtain an approximate solution to
equations~(\ref{eq:mass}) and (\ref{eq:hydrostatic}), and a
mass-radius relationship for the case of a low-mass, slightly
compressed planet.

To zeroth order, $\rho(P)$ does not depend on $P$, so we can integrate
equations~(\ref{eq:mass}) and (\ref{eq:hydrostatic}) to find the zeroth
order solutions:
\begin{equation}
m(r)={4 \over 3} \pi r^3 \rho_0,
\end{equation}
\begin{equation}
\label{eq:analyticP}
P(r) \approx P_c - {2 \over 3} \pi G r^2 \rho_0^2.
\end{equation}
Here,
\begin{equation}
\label{eq:analyticPc}
P_c = {2 \over 3} \pi G R_p^2 \rho_0^2 = {3 G \over {8 \pi}}{M_p^2 \over R_p^4},
\label{eq:pcapprox}
\end{equation}
where $M_p$ is the total mass of the planet, and $R_p$ is the planet's radius.

Now we can use the zeroth order solution for $P(r)$ to write
equation~(\ref{eq:stateepsilon}) to 1st order in the term $f(P(r))$:
\begin{equation}
\rho(P) \approx \rho_0 + f(P(r)=P_c - {2 \over 3} \pi G r^2 \rho_0^2).
\end{equation}
Keeping in mind that the compression is most important in the center, we
will expand this expression for $\rho$ about $r=0$, to get an expression for
the density accurate to first order in $f(P(r))$ and second order in $r/R$:
\begin{equation}
\rho(P_c) \approx \rho_0 + \left[ f(P_c ) - {2 \over 3} \pi G r^2
\rho_0^2 f'(P_c) \right].
\label{eq:rhoapprox}
\end{equation}
If the compressed region is confined to small $r/R$, then we can use this
expression as an approximation for the pressure everywhere in the planet.
We can then substitute this expression into equation~(\ref{eq:mass}) and
integrate to get
\begin{equation}
m(r) \approx {4 \over 3} \pi r^3 \left[ \rho(P_c) - {2 \over 5}
\pi G r^2 \rho_0^2 f'(P_c)\right].
\end{equation}
When we evaluate this equation at $r=R_p$, we find the 
desired mass-radius relationship
\begin{equation}
M_p = {4 \over 3} \pi R^3 \left[\rho(P_c)  - {2 \over 5} \pi G R_p^2 \rho_0^2 f'(P_c) \right].
\label{eq:mvsrapprox}
\end{equation}

The mean density is
\begin{equation}
\label{eq:meandensity}
\bar \rho = \rho(P_c) - \frac{2}{5}\pi G R_p^2 \rho_0^2 f'(P_c).
\end{equation}
We can see that $\rho(P_c) > \bar \rho > \rho_0$, i.e., the mean
density is higher than the zero pressure density but lower than the
density near $r=0$.  Also, for high $P_c$ the radius will decrease
(this is the case for $f(P_c) = cP_c^n$, the
modified-polytropic form in equation~(\ref{eq:modifiedpoly})).

We can evaluate equation~(\ref{eq:meandensity}) to first order in
$f(P(r))$ using equation~(\ref{eq:pcapprox}).  In other words, the
mean density of the planet is approximately
\begin{equation}
\label{eq:analyticrho}
\bar \rho = \rho(P_c) - {3 \over 5} f'(P_c) P_c,
\end{equation}
where $P_c$ is given in equation~(\ref{eq:pcapprox}).

We now apply the above analytic equations 
equations~(\ref{eq:analyticP})--(\ref{eq:analyticrho}) to
our dimensionless equations. We begin by considering our scaled equation of
state as
\begin{equation}
\rho_s (P_s) = 1 + P_s^n.
\end{equation}
We proceed under the assumption that $P_s \ll 1$ and therefore
$P_s^n \ll 1$. For the dimensionless equations we find: the 
scaled pressure
\begin{equation}
P_s(r_s) \approx P_{s,c} - {2 \over 3}\pi r_s^2
\end{equation}
where
\begin{equation}
P_{s,c} = {2 \over 3}\pi R_s^2;
\label{eq:psc}
\end{equation}
the scaled central density 
\begin{equation}
\rho_s (P_{s,c}) \approx 1 +  \left(P_{s,c}^n -  {2 \over 3}\pi r_s^2
n P_{s,c}^{n-1} \right);
\end{equation}
the scaled mass
\begin{equation}
m_s(r_s) \approx {4 \over 3} \pi r_s^3 \left[ 1 +  P_{s,c}^n - {2 \over
5}\pi n r_s^2  P_{s,c}^{n-1}  \right ];
\end{equation}
the desired mass-radius relationship
\begin{equation}
\label{eq:msrs}
\label{eq:massradiusscaled}
M_s \approx {4 \over 3} \pi R_s^3 \left[ 1 + \left(1 - {3 \over 5} n\right) \left({2 \over
3} \pi R_s^2 \right)^n \right];
\end{equation}
and the average density
\begin{equation}
\label{eq:meandensityscaled}
\bar \rho_s = 1 + \left(1 - {3 \over 5} n\right) \left({2 \over
3} \pi R_s^2\right)^n.
\end{equation}

Because we chose $P_s \ll 1$ for this
analytic derivation, we know exactly over what range of parameters
these approximations are valid.  They apply where $P_s \ll 1$, which
is also where $R_s \ll 1$ and $M_s \ll 1$.  The correction term in
equations~(\ref{eq:massradiusscaled}) and
(\ref{eq:meandensityscaled}), $\left[\left(1 - {3 \over 5}
n\right) \left({2 \over 3} \pi R_s^2 \right)^n \right] \ll 1$. The
$R_s \ll 1$ limit therefore shows why the scaled mass-radius relation
depends very weakly on composition: the correction term
in equation~(\ref{eq:massradiusscaled}) is small.

If we consider the scaled mass-radius relationship
equation~(\ref{eq:msrs}) slightly beyond where it is formally valid ($R_s < 1$
instead of $R_s \ll 1$), we find that it is still a reasonable
approximation. While the scaled mass-radius relationship
equation~(\ref{eq:msrs}) is good to within 1\% at $M_s=0.245$
it is good to within 5\% at $M_s=0.36$ (compared to
the numerical solution to the scaled equations).
Even at $R_s \approx 1$, the correction term
is never larger than $(1- {3 \over 5}n)({2 \over 3}\pi)^n$.  Since $n$
ranges typically from 0.5--0.6, this maximum value for the correction term
ranges from about 0.997 to 1.013. In other words, even for $R_s \approx
1$, the correction term in equation~(\ref{eq:msrs}) varies over only a
range of about 0.02 as a function of planet chemistry.

Figure~\ref{fig:mrscaled} compares this analytic mass-radius approximation
to the full numerical solution of the scaled equations, justifying
that our calculations are correct.

\section{Discussion}
\label{sec-discussion}

\subsection{Exoplanet Mass and Radius Observational Uncertainties}

We now discuss the observational uncertainties on the mass and
radius of transiting exoplanets.  We adopt a conservative range for
planet mass and radius fractional uncertainty of 2--10\%. The 10\%
limit is the uncertainty range for typical exoplanet detections,
arising from measurement uncertainty due to the data quality.  Once
discovered, most exoplanets of interest will be followed up with
larger telescopes and/or more observations to refine the planet mass
and radius beyond that determined from the observational discovery.
If measurement uncertainty is not a dominant factor for planet mass
and radius uncertainty, then the stellar mass and radius uncertainty
are. This is because the exoplanet mass and radius are derived from
quantities that involve planet-star mass or radius ratios.  Note that
assuming the stellar noise to be small, the measurement uncertainty
and stellar mass or radius uncertainty add in quadrature.

Our estimate of the 2\% planet mass and radius uncertainty is based on
an optimistic assessment of the 2\% stellar mass and radius
uncertainty likely to be possible for millions of stars in the future.
This kind of high-precision measurement will be enabled with accurate
distances and precise stellar fluxes by the GAIA space mission (ESA;
launch date 2013). In practice the radius can be inferred directly
from the stellar fluxes and distances; in principle a precise radius
is limited by the correction from a measured stellar flux to the
star's bolometric flux (D. Sasselov, 2006 private communication).  In
contrast to our 2\% best case scenario, current typical stellar mass
and radius uncertainties are on the order of 5--10\% \citep{ford1999,
cody2002, fisc2005}. These stellar masses and radii are derived from
interior and evolution model fits to observed stellar spectra.  
\citet{sozz2007} show that a more precise stellar radius (and hence
planet radius) can be derived using stellar evolution models in in the
$a/R{_*}$ vs. $T_{eff}$ parameter space instead of in the usual
log~$g$ vs. $T_{eff}$ parameter space.  This is partly because $a/R_*$
(a measure of stellar density) can be determined with high precision
from the planet transit light curve for planet's with
zero orbital eccentricity \citep{seag2003}.  Provided the
photometry is good enough, few percent uncertainty in star and planet
radii may become routine.  We note that a
different technique, interferometry, can measure stellar radii
directly, but the current uncertainties are much higher than 2\% and
the technique is limited to a small number of nearby stars.

The mass determination of low-mass planets by ground-based radial
velocity techniques require hundreds of observations.  Mass
determination for many exoplanets may therefore be inhibited
\citep{lovi2006} until a number of dedicated ground-based telescopes
are available.  Given such limitations of telescope time and current
technology, the optimistic future exoplanet mass uncertainty range is
likely closer to 5--10\% for the low-mass solid planets of interest
($< 20 M_{\oplus}$).  With current technology and due to the faintness
of the host stars, many low-mass exoplanets discovered from transit
surveys (e.g., COROT or Kepler) will not have measured masses at all.
As an example of what it would take to detect an Earth-mass planet in
a 50-day orbit about one of the brightest sun-like stars: one
eight~meter diameter telescope dedicated to five bright stars
monitored over five years (R.~P. Butler 2006, private
communication). Earth-mass planets more distant from their host stars,
such as Earth-like orbits about sun-like stars cannot be detected with
any current technology.

\subsection{Possible Exoplanet Compositional Distinctions}

Given the observational uncertainties, what compositional distinctions
among exoplanets can we make?  With an upper limit of  20\% 
on the planet mass uncertainty we
will be able to say robustly that the planet is predominantly composed
of solids or if it instead has a significant gas envelope as do
Uranus and Neptune.

The following discussion assumes we can ascertain that the planet
has no substantial atmosphere or envelope that would contribute
to the planet radius (c.f. Adams et al., in prep).

With a 10\% uncertainty we may be able to comment on the presence of a
large amount of water or iron---if the planet fortuitously has a very
low density or a very high density within the radius range for solid
planets.

A $\sim 5\%$ uncertainty will allow us to distinguish among planets
composed predominantly of water-ice, predominantly of silicates, and
predominantly of iron (sections~\ref{sec-homoplanets} and
\ref{sec-diffplanets}).  These planets are relatively well separated
on the mass-radius diagram (Figure~\ref{fig:mrplanets}). Their
separation is primarily due to the low density of water ice, the
intermediate density of silicates, and the high density of Fe.  Even
with 5\% exoplanet mass and radius uncertainties, it is not possible
to identify the detailed composition such as the fraction of different
material in the core and differentiated layers.

Identification of water planets is possible with $\sim 5$\% planet
mass and radius uncertainty if the planet has more than 25\% water ice
by mass and plausible iron-to-silicate ratios. With the same
$\sim$~5\% uncertainty, water planets with 50\% water by mass with any
iron-to-silicate ratio can be identified. Water exoplanets should
exist; in our own solar system Jupiter's satellite Ganymede is 45\%
water ice by mass. Water ice planets should be the easiest of the
solid exoplanets to detect observationally because of their large
radius for a given planet mass---they could be as large as $3
R_{\oplus}$ for $M_p = 20 M_{\oplus}$. We note that \citet{vale2007a}
came to the same conclusion for the one planet mass they explored
water compositions of, the 7.5 $M_{\oplus}$ Gl 876d. A detection of a
low-density water-ice planet orbiting far interior to the expected
snow line in the disk where the planet formed would be strong evidence
for planet migration.
 
With 2\% uncertainty in planet mass and radius we would be able to
determine not only the basic composition of any iron/silicate/water
planet, but constrain the relative fraction of each material.
However, even with 2\% uncertainty in planet mass and radius many
degeneracies in planet composition remain for a planet of the same
mass and radius. For example, carbon planets with silicate mantles and
iron cores are indistinguishable from silicate planets with small iron
cores, or with planets that have a small ($\sim 10$\% by mass)
fraction of water on top of a silicate mantle and iron core. As a
second example, planets with deep water oceans (even 100~km deep) on
any kind of planet are not identifiable with even 2\% fractional
uncertainty in the radius.  This is because a liquid water ocean
contributes only a small amount to the overall planet radius compared
to a pure water ice layer. For a third example, see
Figure~\ref{fig:rhocurves}d and the accompanying discussion. See also
\citet{vale2007b} for a detailed discussion of degeneracies in
composition for iron-silicate-water planets of 1--10 $M_{\oplus}$.

Observational uncertainties in planet mass and radius are unlikely to
be better than a few percent in the next decade. We therefore argue that
detailed exoplanet interior models are not needed to infer exoplanet
bulk composition.  The analytical form derived in this paper
(equation~(\ref{eq:scaledmrapprox})), with scaling relations provided
from model curves (e.g., in Table~4), should be sufficient for the
near future.

Exoplanet atmosphere measurements may inform us about the planet
interior, removing some of the degeneracies in a given solid
exoplanet mass and radius relationship. Therefore, despite our
conclusions that detailed interior planet models are not needed to
determine exoplanet bulk composition, we point out that detailed
interior models are required to understand planetary atmospheres.  In
turn, observations of exoplanet atmospheres may help us infer more
about the exoplanet interior composition. Atmosphere measurements will
also help us to identify interesting features on planets, such as the
presence of a deep liquid ocean.  A saturated water vapor atmosphere
would be a better indicator than an exoplanet mass and radius,
considering we cannot identify deep water oceans with mass-radius
relationships.

\section{Summary and Conclusions}

We have modeled cold solid planets of a variety of compositions
including iron, silicates, water, and carbon compounds. The main
conclusions of this work are:

1. All solid planets approximately follow the scaled mass-radius
relationship $\log_{10} R_s = k_1 + 1/3 \log_{10}(M_s) - k_2
M_s^{k_3}$ for up to $M_p \simeq 4 M_{s}$.  This relationship can be
scaled to physical units by the values $m_1$ and $r_1$ given in
Table~4. The corresponding planet mass, in physical units, to which
the above mass-radius equation is valid ranges from $M=17 M_{\oplus}$
to $M_p = 40 M_{\oplus}$, depending on the material (see the $m_1$ values
in Table 4.)

2. There is no simple power law for the mass-radius relation; the
mass-radius curve slope changes even within a relatively narrow mass
range.

3.  We can use the same formalism described in the above equation to
summarize the mass-radius relationships we computed for differentiated
planets, using the scaling parameters listed in Table~4.  Given the
uncertainties in the equations of state and the best expected future
observations, this simple, handy approximation supplies enough detail
and accuracy to interpret any forecasted observations.

4. Highly detailed interior planet models are not needed to
infer a solid exoplanet's bulk composition from its mass and radius.
This is because
the temperature structure and phase changes have little
impact on the total planet mass and radius.

a. Low-pressure phase changes (at $< 10$ GPa) are not important for a
planet's radius because for plausible planet compositions most of the
mass is at high pressure. For example, 97\% of the planet's mass is at
high pressures ($>$3 GPa). For high pressure phase changes we expect
the associated correction to the EOS (and hence derived planet radii)
to be small because at high pressure the importance of chemical
bonding patterns to the EOS drops.

b. Temperature can be approximated as a thermal pressure term. This
thermal pressure causes a decrease in density of on order three
percent or less at relevant temperatures and pressures.  A change in
average density of a planet translates into a smaller change in the
planet radius, because $R_p \sim \bar{\rho_p}^{-1/3}$. Conceptually,
At low pressures ($\lesssim$ 10 GPa) in the outer planetary layers,
the crystal lattice structure dominates the material's density and the
thermal vibration contribution to the density are small in comparison.
At high pressures the thermal pressure contribution to the EOS is
small because close-packed nature of the materials prevents structural
changes from thermal pressure contributions.  

5. We identified several interesting properties of exoplanets.

a. Planets are not likely to be found that have radii smaller than a
pure Fe planet. Fe is the most dense element out of which planets are
expected to form. While not a new conclusion, this point
is useful to keep in mind for designing and interpreting radius observations
of exoplanets.

b. Planets above the H$_2$O curve must have a significant H/He
envelope. We can therefore easily distinguish between
exoplanets with significant H/He envelopes and those without, as is
the case for GJ~436b. We therefore define a ``super Earth'' to be a
solid planet with no significant gas envelope, regardless of its mass.

c. Because of their unique position on the mass-radius diagram, H$_2$O
planets with more than 25\% water ice by mass can be identified with
approximately 5\% fractional uncertainty in $M_p$ and $R_p$.  (A
similar conclusion was also found by \citet{vale2007a}; that if the 7.5
$M_{\oplus}$ planet they modeled has a large water content it would be
identified by a large radius.)  Discovering a water planet orbiting at
a small semi-major axis would serve as strong evidence for planet
migration, since presumably water planets form more efficiently far
from their host stars, beyond the ice line.  This point is valid
provided the planet is not a carbon planet and provided there is no
substantial contribution to the planet radius from
an atmosphere or envelope.

d. Even with planet mass and radius measurement uncertainties better
than 1\%, planets of different interior composition can have the same
total mass and radius. In other words, different mass fractions
of iron cores, silicate mantles, and water outer layers can have
the same total radius for a planet of the same mass.

e. Carbon planets, if they exist, have mass-radius relationships that
overlap with mass-radius relationships of non-carbon planets (i.e.,
water and silicate planets).  This is because the zero-pressure
density of graphite is similar to that of H$_2$O and the zero-pressure
density of SiC is similar to that of MgSiO$_3$.  Exoplanet atmospheres
of transiting exoplanets will have to be observed and studied to
distinguish between carbon planets and water/silicate planets.

6. We conclude that, while  detailed interior structure models
are needed to understand the atmosphere formation and evolution, 
detailed interior structure models are not needed to infer
bulk composition from exoplanet mass and radius measurements.

\acknowledgements
We dedicate this study of mass-radius relationships
to the memory of our co-author Cathy Hier-Majumder. It is with great
sadness that we acknowledge her untimely death.  We thank Rus Hemley,
Y. Fei, and Dan Shim for extremely useful discussions about high pressure
physics and equations of state. We thank Mercedes Lopez-Moralez, Dimitar
Sasselov, Jim Elliot for useful discussions about observational
uncertainties. We thank Diana Valencia for providing a more detailed
version of Figure~4 in \citet{vale2006}, as well as for interesting
discussions about unpublished work. This work was supported by the
Carnegie Institution of Washington, the NASA Astrobiology
Institute, and the Massachusetts Institute of Technology.

\bibliography{planets}

\clearpage

\begin{table}

\begin{tabular}{l l l l l l l}

\hline

Atom or Compound & $K_0$ (GPa) & $K'_0$ & $\rho_0$ (Mg m$^{-3}$) & Fit &log$_{10} P_{V/T}$ (GPa) & Ref. \\
\hline
\hline
C (graphite)        &  33.8$\pm$3 & 8.9$\pm$1.0 & 2.25 & BME &  11.75 & 1,2\\
Fe ($\alpha$)   &  162.5 $\pm$5 & 5.5 $\pm$0.8 & 7.86 & BME &--& 1,3 \\
Fe ($\varepsilon$)   &  156.2 $\pm$1.8 & 6.08$\pm$0.12 & 8.30 &V & 13.32 & 4 \\
FeS                 & 35$\pm$4 & 5$\pm$2 & 4.77 & BME &13.23 & 1,5\\
H$_2$O (ice VII)    &  23.7$\pm$0.9  & 4.15$\pm$0.07 & 1.46 & BME & -- & 6\\
H$_2$O (liquid)$^a$   & 2.28 & -- & -- & -- & -- & 7 \\
MgO                 &  177.0$\pm$4 & 4.0$\pm$0.1  & 3.56 & BME & 12.8 & 8 \\
MgSiO$_3$ (en)      &  125   & 5$^b$  & 3.22  & BME &-- & 1,9 \\
MgSiO$_3$ (pv)      &  247$\pm$4   & 3.97$^c$  & 4.10  & BME4 & 13.13 & 10 \\
(Mg$_{0.88}$,Fe$_{0.12}$)SiO$_3$ (pv)      &  266$\pm$6   & 3.9$\pm$0.04  & 4.26  & BME & 12.74 & 1,11 \\
SiC                 &  227$\pm$3 & 4.1$\pm$0.1    & 3.22       & BME & 11.4 & 1,12  \\
\hline

\end{tabular}

\caption{Parameters for the Vinet (V) or Birch-Murnagham (BME) EOS
fits.  References: (1) \citet{ahre1995}, (2) \citet{hanf1989}; (3)
\citet{taka1968}; (4) \citet{ande2001}; (5) \citet{king1982} (6)
\citet{heml1987}; (7)\citet{hali2003}; (8) \citet{duff1995}; (9)
\citet{olin1977}; (10) \citet{kark2000}; (11) \citet{knit1987}; (12)
\citet{alex1989}.  a) seawater at 12$^{\circ}$K. b) $K'_0$ values are assumed. c) A fourth
order BME fit was used with $K''_0$ = -0.016/GPa.}

\end{table}

\begin{table}

\begin{tabular}{l l l}

\hline
$V$ (cm$^3$/mol) & $\rho$ (kg m$^{-3}$)  & P (GPa) \\
\hline
10.998300 & 1.636617 & 2.320 \\
10.429585 & 1.725860 &4.155 \\
9.880818 & 1.821712 &6.664 \\
9.351623 & 1.924800 & 9.823 \\
8.350443 &  2.155574 & 18.791 \\
7.878082 & 2.284820 & 25.361 \\
7.423411 & 2.424761 & 33.744 \\
6.986806 & 2.576285 & 44.314 \\
6.567891 & 2.740606 & 56.970 \\
6.165913 & 2.919275 & 74.188 \\
5.780497 & 3.113919 & 94.406 \\
5.411641 & 3.326163 & 126.815\\
5.164734 & 3.485175 & 155.924 \\
4.654734 & 3.867031 & 240.696 \\
4.195170 & 4.290649 & 351.114 \\
3.780772 & 4.760933 & 498.660 \\
3.407399 & 5.282621 & 691.938 \\
3.070912 & 5.861450 & 937.585 \\
2.767547 & 6.503954 & 1260.182 \\
2.494293 & 7.216474 & 1673.049 \\
2.248138 & 8.006625 & 2188.301 \\
2.026072 & 8.884186 & 2853.712 \\
1.825836 & 9.858497 & 3691.387 \\
1.645548 & 10.938603 & 4737.211 \\
1.483327 & 12.134882 & 6040.611 \\
1.336538 & 13.467635 & 7686.171 \\
\hline
\end{tabular}
\caption{Density functional theory (DFT) EOS for water ice VIII and X. DFT predicts a gradual transition between the two phases.}
\end{table}

\begin{table}

\begin{tabular}{l l l l}

\hline Material & $\rho_0$ [kg m$^{-3}$] & $c$ [kg m$^{-3}$ Pa$^{-n}$] & $n$ \\ \hline

Fe($\alpha$) & 8300.00 & 0.00349 & 0.528 \\
MgSiO$_3$ (perovskite) & 4100.00 & 0.00161 & 0.541 \\
(Mg,Fe)SiO$_3$ &  4260.00 & 0.00127 & 0.549 \\
H$_2$O &  1460.00 & 0.00311 & 0.513 \\
C (graphite) & 2250.00 & 0.00350 & 0.514 \\
SiC & 3220.00 & 0.00172 & 0.537 \\
\hline

\end{tabular}

\caption{Fits to the merged Vinet/BME and TFD EOS of the form $\rho
(P) = \rho_0 + c P^n$. These fits are valid for the pressure range $P
< 10^{16}$ Pa.}

\end{table}

\clearpage
{\rotate
\begin{table}

\begin{tabular}{l l l l l l l}

\hline Material & $m_1$ [$M_{\oplus}$] & $r_1$ [$R_{\oplus}$] & $P_1$ [GPa] &$k_1$ & $k_2$ & $k_3$ \\ \hline

Fe($\alpha$) [modified-polytropic EOS]           & 5.80 & 2.52 & 1192 & -0.209490 & 0.0804 & 0.394\\
MgSiO$_3$ (perovskite) [modified-polytropic EOS]& 10.55 & 3.90 & 693 & -0.209594 & 0.0799 & 0.413\\
H$_2$O (ice) [modified-polytropic EOS]                & 5.52 & 4.43 & 114 & -0.209396 & 0.0807 & 0.375\\
\hline
Fe($\alpha$)           & 4.34 & 2.23 &  &  &  & \\
MgSiO$_3$ (perovskite) & 7.38 & 3.58 &  &  &  & \\
H$_2$O  (ice)               & 8.16 & 4.73 &  &  &  & \\
Fe(0.675)/MgSiO$_3$(0.325) & 6.41& 3.19 &  &  &  & \\
Fe(0.3)/MgSiO$_3$(0.7) & 6.41 & 2.84 &  &  &  & \\
Fe(0.225)/MgSiO$_3$(0.525)/H$_2$O(0.25) & 6.41 & 3.63 &  &  &  & \\
Fe(0.065)/MgSiO$_3$(0.485)/H$_2$O(0.45) & 6.88 & 4.02 &  &  &  & \\
Fe(0.03)/MgSiO$_3$(0.22)/H$_2$O(0.75) & 7.63 & 4.42 &  &  &  & \\
\hline

\end{tabular}

\caption{Conversion factors for the scaling relationships for
equation~(\ref{eq:scaledmrapprox}).
The conversion factors give the physical values from the scaled
parameters mass, radius, pressure, and density. See
equations~(\ref{eq:scalerho}) through (\ref{eq:scalem}). The first
three rows of this table additionally give parameters for mass-radius
relationships computed from the modified-polytropic-EOS; these include
the $k_i$ values for a fit to equation~(\ref{eq:scaledmrapprox}) valid
for $M_s > 4$.}
\end{table}
}

\clearpage

\begin{figure}
\begin{center}
\includegraphics[width=14cm, height=14cm]{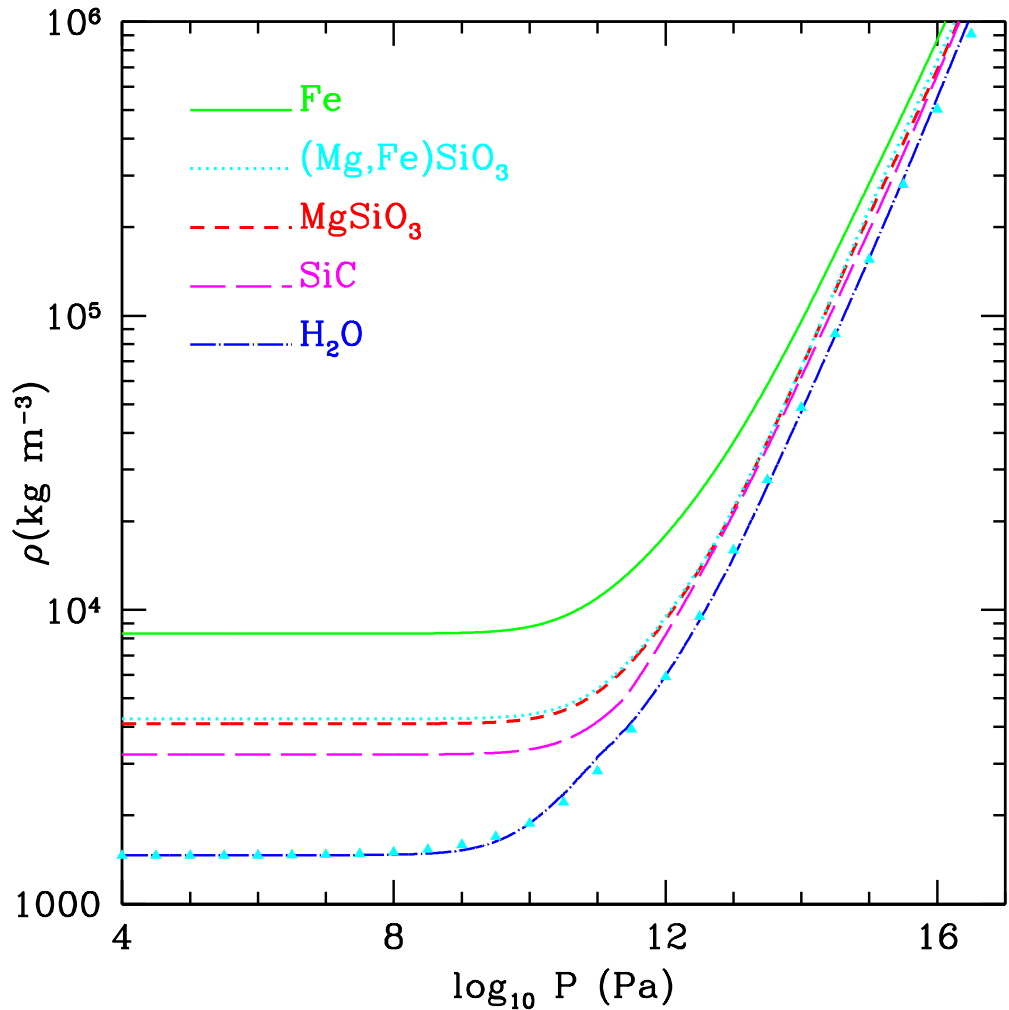}
\end{center}
\caption{Equations of state for five different materials used in this
study.  Each EOS data set was derived by combining a fit to
experimental results and the TFD limit at high pressure. For water ice
we used density functional theory as a bridge between experiment and
the TFD theory. The EOSs are all reasonably well approximated by a
polytropic-like expression $\rho (P) = \rho_0 + c P^n$, where $\rho_0$
is the zero-pressure density and $c$ and $n$ are constants. Table~2
lists these constants for some materials.  The cyan triangles show one
such fit for the H$_2$O EOS.
\label{fig:EOSsingle} }
\end{figure}

\begin{figure}
\begin{center}
\includegraphics[width=14cm, height=14cm]{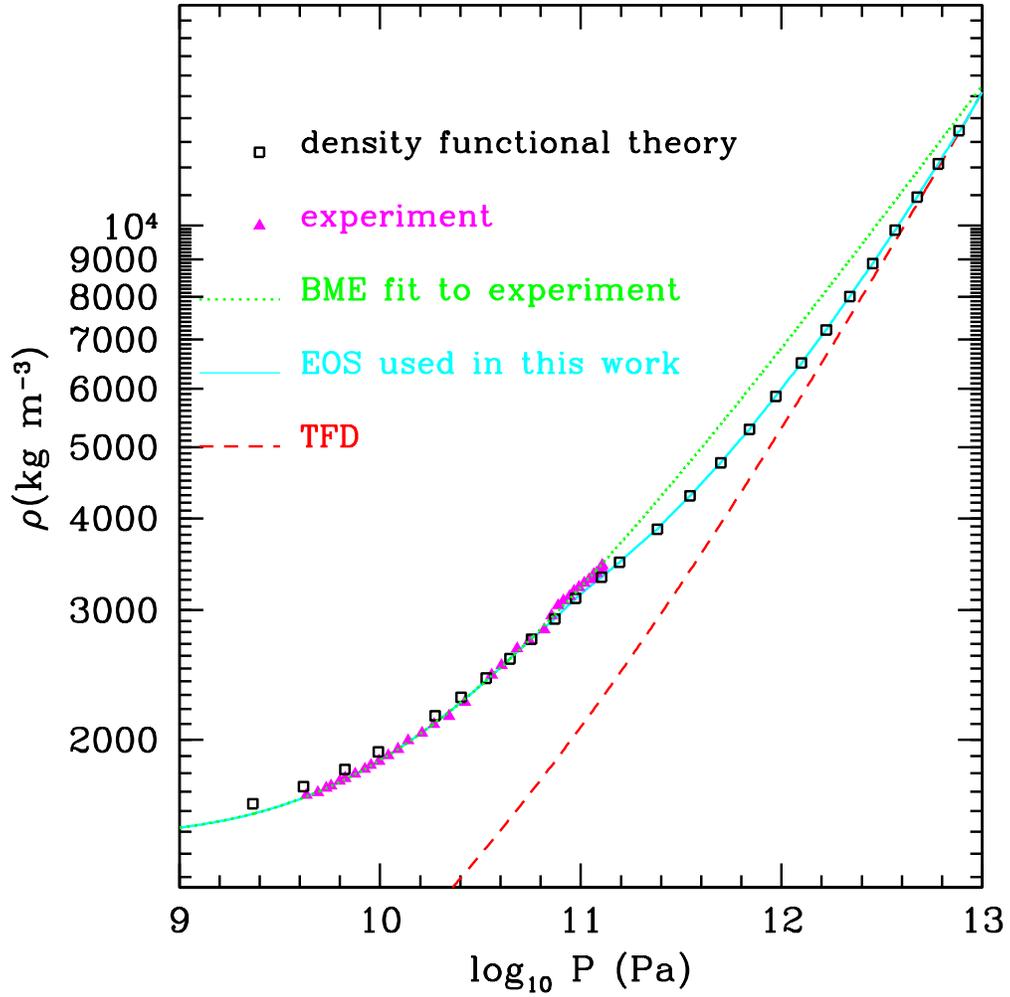}
\end{center}
\caption{The H$_2$O EOS used in this study (solid cyan curve).  We use
the BME fit to the experimental data \citep[][magenta
triangles]{heml1987} up to $P = 44.3$~GPa.  At this pressure we switch
to the density functional EOS (squares). At $P=7686$~GPa we switch to
the TFD EOS (red dashed curve).
\label{fig:h2oEOS} }
\end{figure}

\clearpage

\begin{figure}
\begin{center}
\includegraphics[width=14cm, height=14cm]{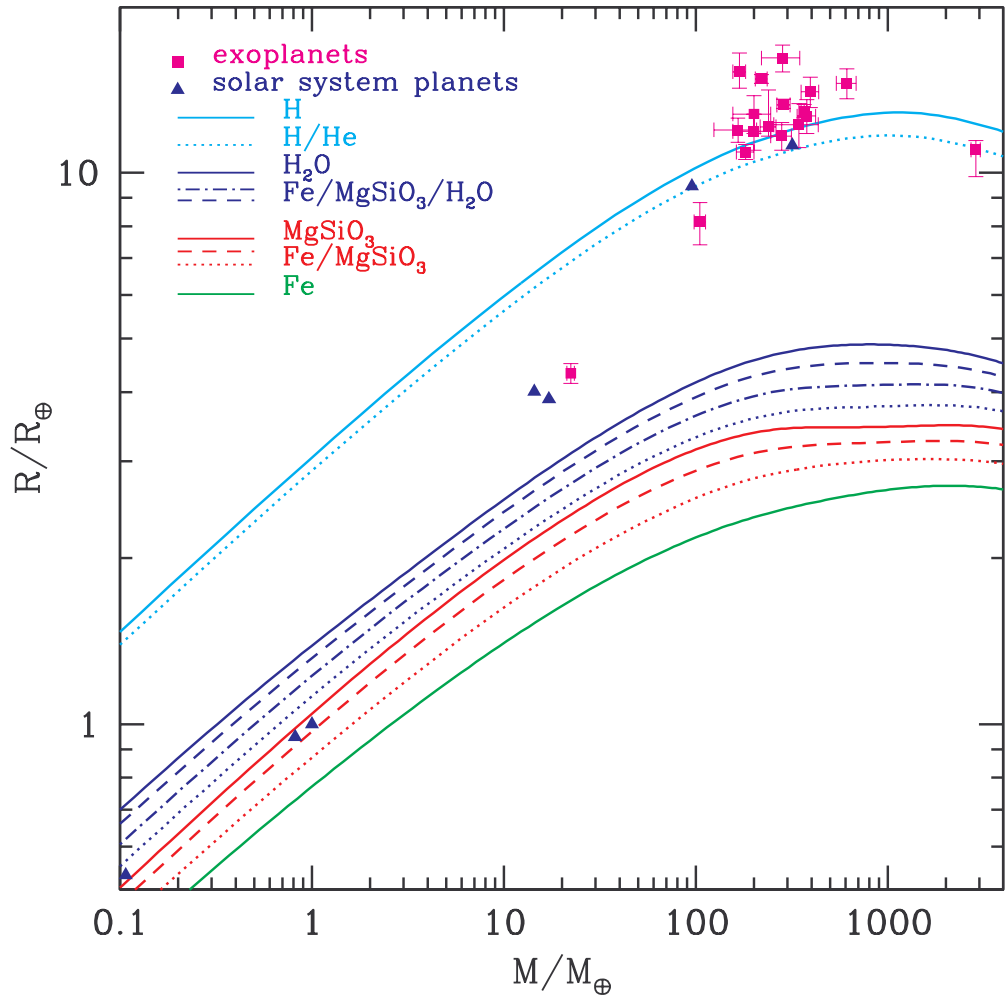}
\end{center}
\caption{Mass-radius relationships for solid planets.  The solid lines
are homogeneous planets. From top to bottom the homogeneous planets
are: hydrogen (cyan solid line); a hydrogen-helium mixture with 25\%
helium by mass (cyan dotted line); water ice (blue solid line);
silicate (MgSiO$_3$ perovskite; red solid line); and iron (Fe
($\epsilon$); green solid line). The non-solid lines are
differentiated planets.  The red dashed line is for silicate planets
with 32.5\% by mass iron cores and 67.5\% silicate mantles (similar to
Earth) and the red dotted line is for silicate planets with 70\% by
mass iron core and 30\% silicate mantles (similar to Mercury). The
blue dashed line is for water planets with 75\% water ice, a 22\%
silicate shell and a 3\% iron core; the blue dot-dashed line is for
water planets with 45\% water ice, a 48.5\% silicate shell and a 6.5\%
iron core (similar to Ganymede); the blue dotted line is for water
planets with 25\% water ice, a 52.5\% silicate shell and a 22.5\% iron
core. The blue triangles are solar system planets: from left to right
Mars, Venus, Earth, Uranus, Neptune, Saturn, and Jupiter. The magenta
squares denote the transiting exoplanets, including HD~149026b at 8.14
$R_{\oplus}$ and GJ~436b at 3.95 $R_{\oplus}$.  Note that electron
degeneracy pressure becomes important at high mass, causing the planet
radius to become constant and even decrease for increasing mass.
\label{fig:mrplanets} }
\end{figure}

\begin{figure}
\begin{center}
\includegraphics[width=14cm, height=14cm]{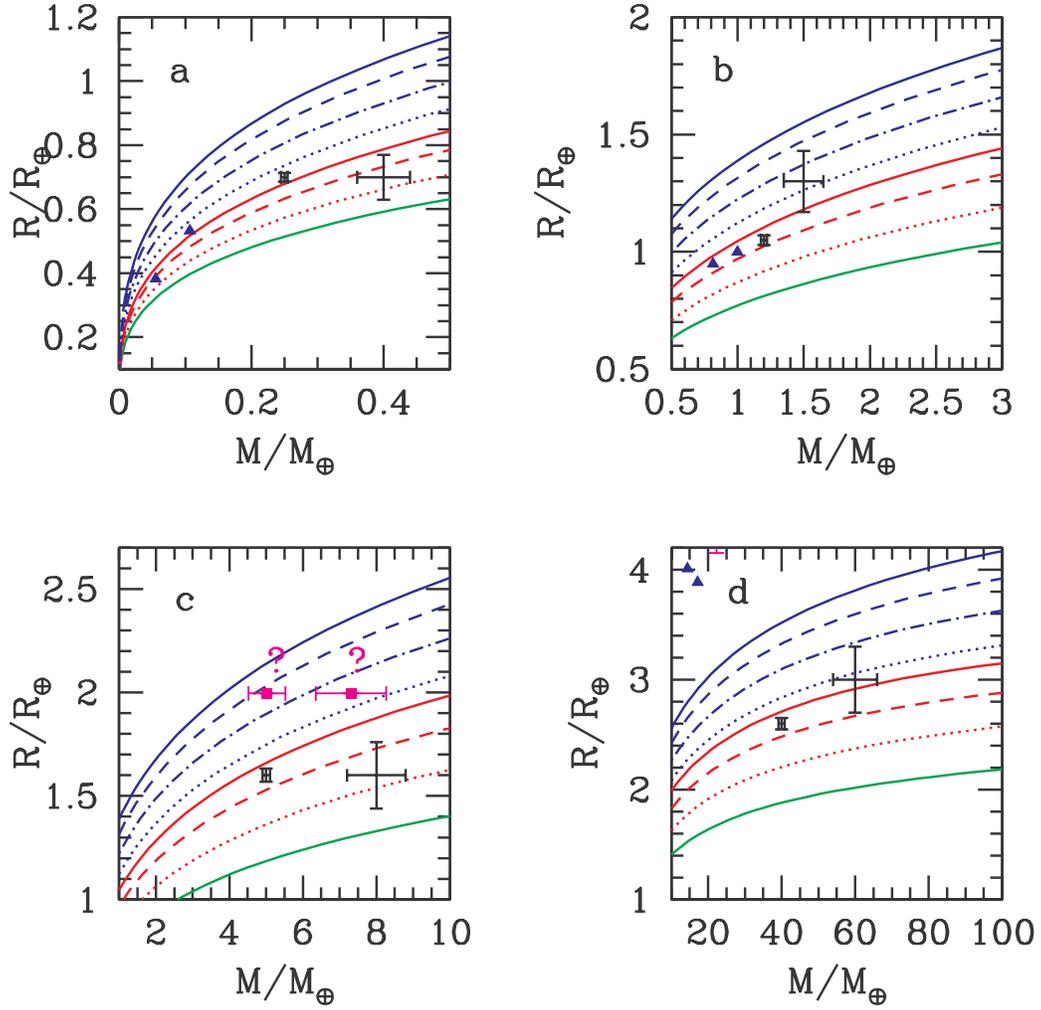}
\end{center}
\caption{Mass-radius relationships for planets with $R_p < 4
R_{\oplus}$. The curves are as in Figure~\ref{fig:mrplanets}: blue are
for water ice planets, red are for silicate planets, and green is for
pure iron planets. Black error bars are shown for 2\% and 10\% uncertainty
in planet mass and radius.  Each panel shows a different mass range.
The terrestrial-mass solar system planets are shown with blue
triangles. The exoplanets Gl~876d and Gl~531c are shown with a
magenta square in panel c; although the radii are not known the are
shown to represent known low-mass exoplanets. The exoplanet GJ~436b is
shown in panel~d.
\label{fig:mrplanetszoom} }
\end{figure}

\begin{figure}
\begin{center}
\includegraphics[width=14cm, height=14cm]{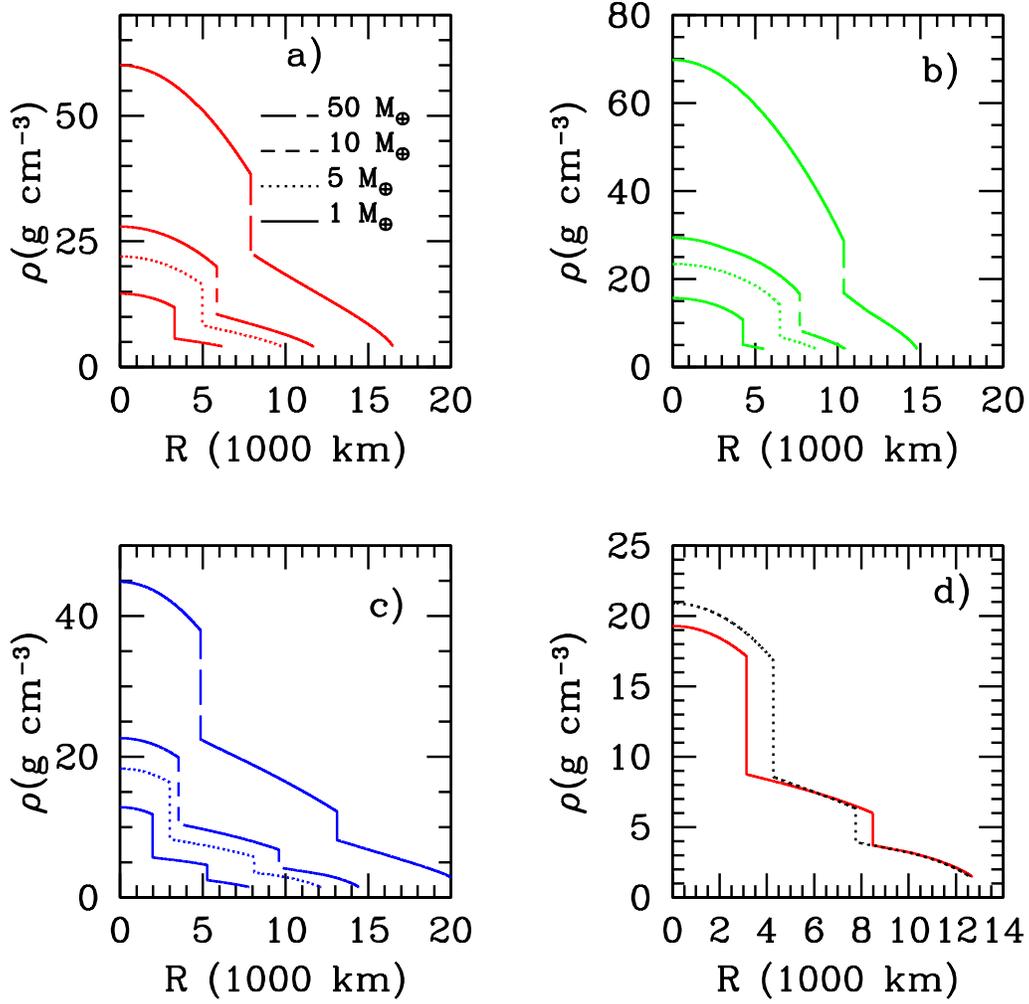}
\end{center}
\caption{Interior structure of solid exoplanets. From top to bottom
the curves in panels a, b, and c are for planets with $M_p=$ 50, 10,
5, and 1 $M_{\oplus}$ respectively.  Panel a: silicate planets with a
32.5\% by mass Fe core and a 67.5\% MgSiO$_3$ mantle. Panel b: as in
panel a but for planets with a 70\% Fe core and 30\% silicate
mantle. Panel c: interior structure for water planets with 6.5\% Fe
core, 48.5\% MgSiO$_3$ shell, and 45\% outer water ice layer. Panel d:
interior model for two different water exoplanets with the same planet
mass and radius: $M_p = 6.0 M_{\oplus}$ and $R_p = 2.0
M_{\oplus}$. The solid curve is for a model with layers in percentages
by mass of Fe/MgSiO$_3$/H$_2$O of: 17/33/50 (similar to the
composition of the water planet in \citep{lege2004}) and the dotted
line for 6.5/48.5/45 (similar to the composition of Ganymede).
\label{fig:rhocurves}}
\end{figure}

\begin{figure}
\begin{center}
\includegraphics[width=14cm, height=14cm]{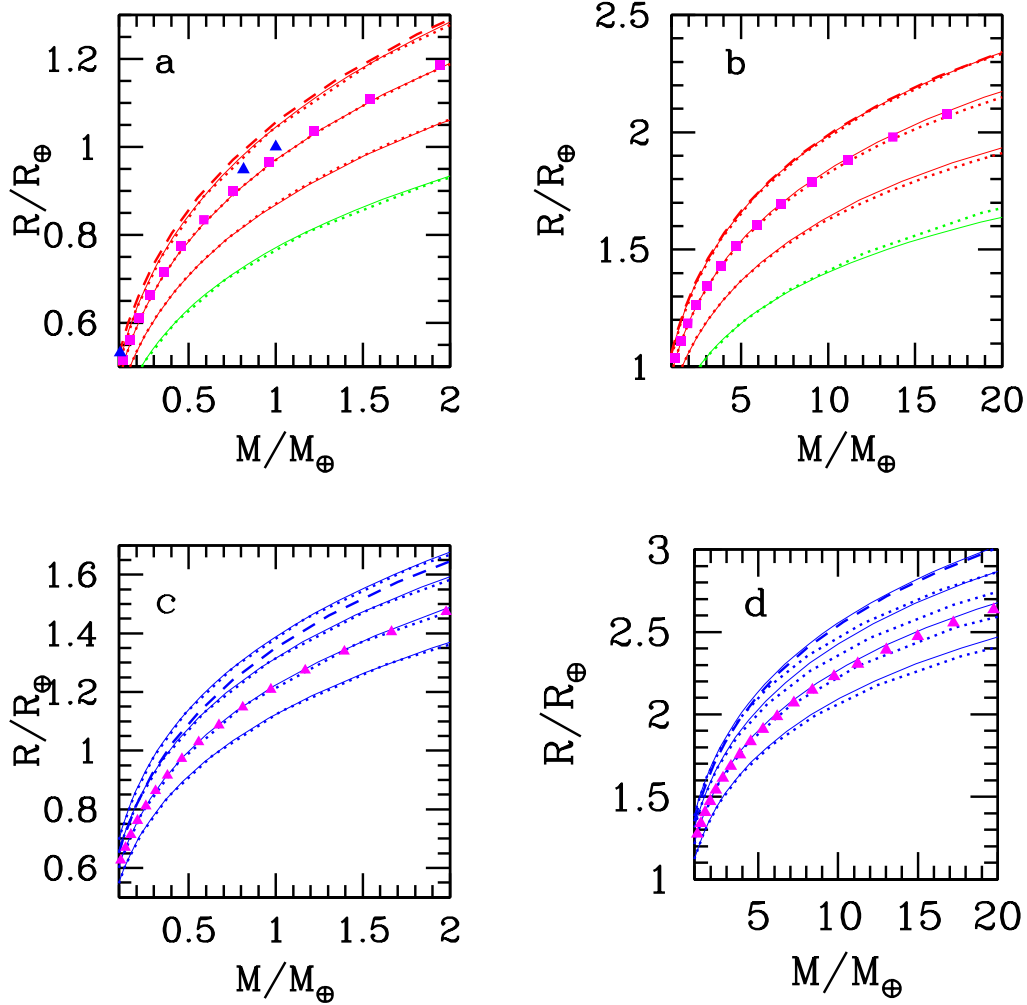}
\end{center}
\caption{Mass-radius relationships of solid planets to illustrate
phase effects. Panel a and b: from top to bottom the solid curves are
for: pure MgSiO$_3$ perovskite planets; MgSiO$_3$ planets with a
32.5\% by mass Fe($\epsilon$) core; MgSiO$_3$ planets with a 70\% Fe
core; and pure Fe planets. The dotted lines just above or beneath each
of the solid curves is the planet mass-radius relationship for the BME
or Vinet EOS alone without the TFD EOS. The top dashed line shows the
pure MgSiO$_3$ perovskite planet with a phase change to MgSiO$_3$
enstatite at pressures less than 10 GPa.  The squares show ``super
Earths'' composed of a 32.5 \% FeS core by mass and a 67.5\% mantle;
the mantle itself composed of 90\% (Mg,Fe)SiO$_3$ by mass mixed with
10\% MgO.  Panels c and d: from top to bottom the solid lines are
water ice planets with layers in percentages by mass of
Fe/MgSiO$_3$/H$_2$O of: 0/0/100; 3/22/75; 6.5/48.5/45, 22.5/52.5/25.
The dotted lines beneath each of the solid curves is the planet
mass-radius relationship for the BME form of H$_2$O ice, without using
our adopted H$_2$O EOS (but using our adopted EOSs for MgSiO$_3$ and
Fe). The dashed line shows the pure water ice planet with liquid water
below 10 GPa. The magenta triangles show the mass-radius relationship
for water planets with 17/33/50, illustrating a degeneracy with the
6.5/48.5/45 water planet (see also Figure~\ref{fig:rhocurves}d).
\label{fig:phase}}
\end{figure}

\begin{figure}
\begin{center}
\includegraphics[width=14cm, height=14cm]{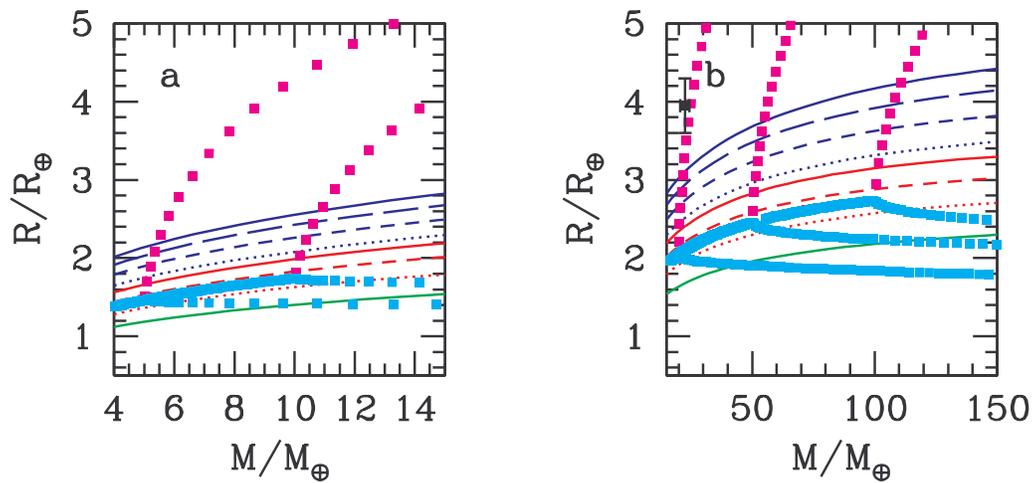}
\end{center}
\caption{The effect of a H/He gas envelope on planet radius, showing
that only a small mass of H/He gas contributes a large factor to the
planet radius.  The lines in this Figure are the same as those in
Figure~\ref{fig:mrplanets}. The magenta squares are planets with a
fixed core mass composed of a mixture of 30\% Fe by mass and 70\%
MgSiO$_3$. The cyan squares show the core mass and radius only. Panel
a shows planets with a fixed core mass of $5 M_{\oplus}$ and $10
M_{\oplus}$ (from left to right).  Panel b shows planets with a fixed
core mass of $20 M_{\oplus}$, $50 M_{\oplus}$ and $100 M_{\oplus}$
(from left to right). These H/He planet radii are underestimates
because they are for zero temperature; temperature would make
the planets larger for a given mass (i.e, move the squares
up and left in this figure). 
\label{fig:gasplanets} }
\end{figure}

\begin{figure}
\begin{center}
\includegraphics[width=14cm, height=14cm]{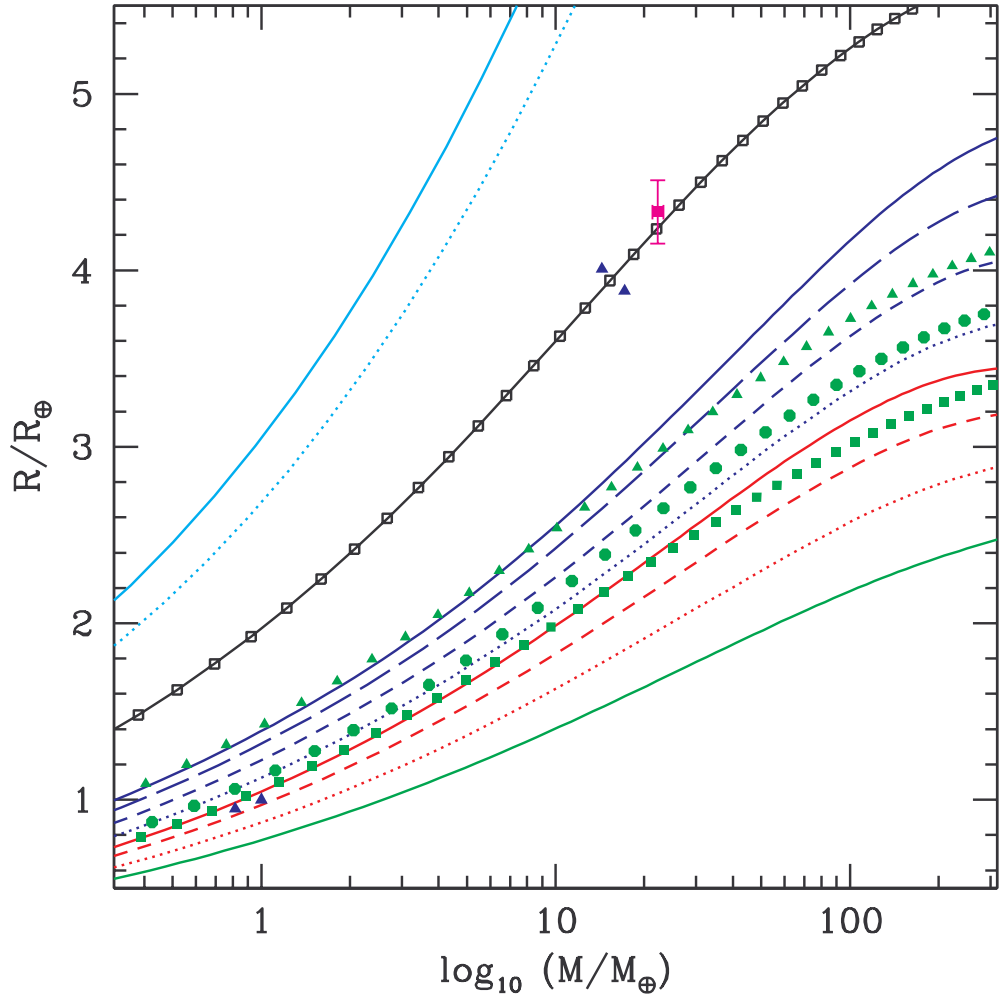}
\end{center}
\caption{Mass-radius relationship for carbon planets and helium
planets.  The curves are the same as those in
Figure~\ref{fig:mrplanets}.  The carbon planet mass-radius
relationships are shown for: carbon monoxide planets (green triangles);
graphite planets with 30\% Fe cores by mass and 70\% graphite mantles
(circles); and SiC planets with 30\% Fe cores by mass and 70\% SiC
mantles (squares). Pure cold He planet mass-radius relationships are
shown by open squares connected by a solid line. The solar system
planets are shown by the blue triangles.
\label{fig:carbonplanets} }
\end{figure}

\begin{figure}
\begin{center}
\includegraphics[width=14cm, height=14cm]{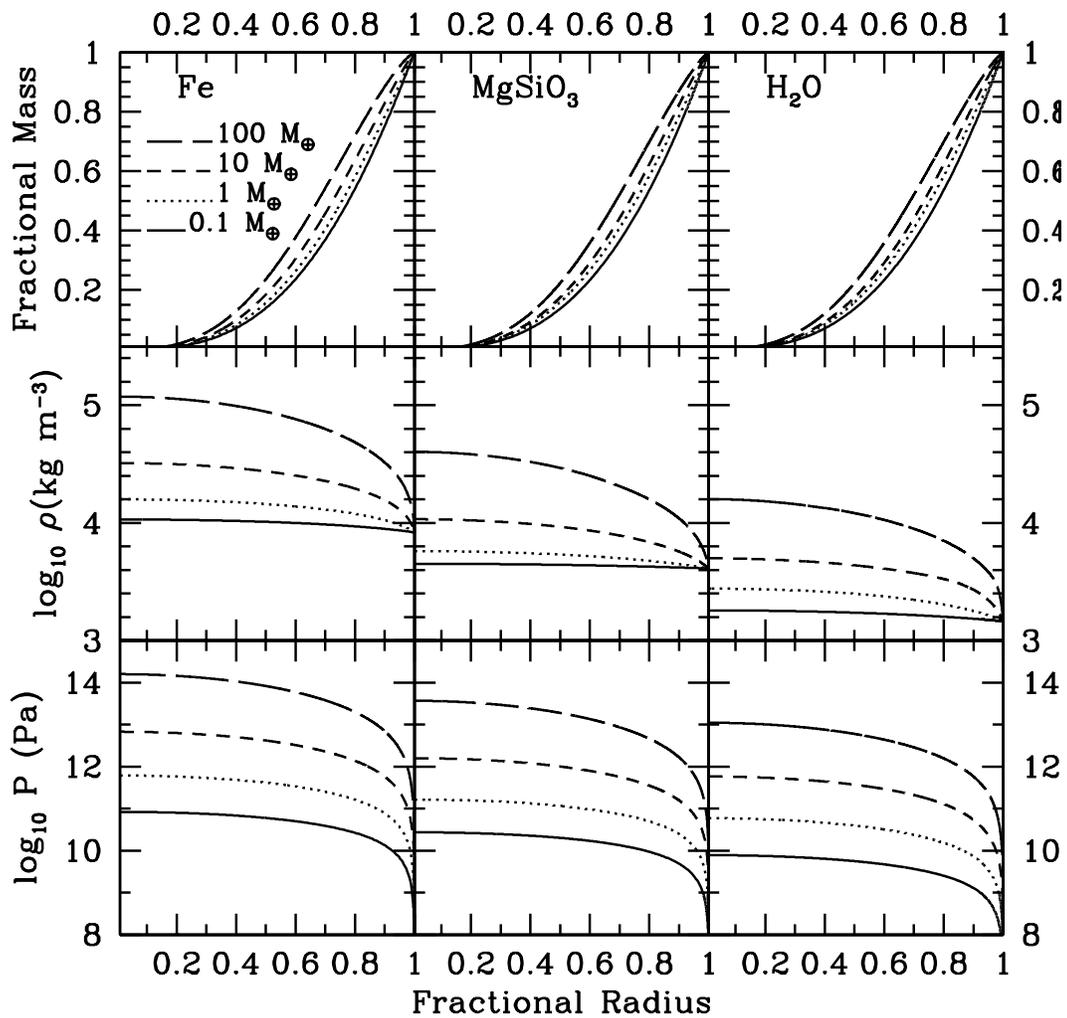}
\end{center}
\caption{Interior structure of homogeneous planets of 0.1, 1, 10, and
100 $M_{\oplus}$. The top row shows the fractional mass as a function
of fractional radius. The middle row shows the density as a function
of fractional radius. The bottom row shows the pressure as a function
of fractional radius.
\label{fig:mrfn}}
\end{figure}

\begin{figure}
\begin{center}
\includegraphics[width=14cm, height=14cm]{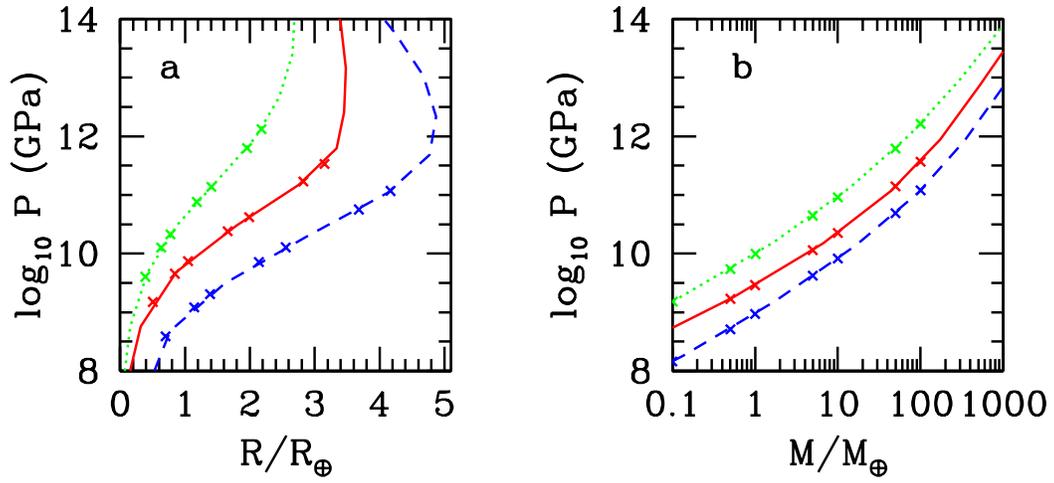}
\end{center}
\caption{The pressure level that contains 97\% of the planet radius or
mass.  Panel a: the pressure at which the planet is 97\% of its total
size, as a function of total planet radius. Panel b: The pressure at
which the planet contains 97\% of its total mass, as a function of
planet mass. The blue dashed curve is for pure water ice planets, the
red solid curve is for pure silicate planet, and the green dotted
curve is for pure iron planets, and
H$_2$O ice VII.
\label{fig:fracmassradius}}
\end{figure}

\begin{figure}
\begin{center}
\includegraphics[width=14cm, height=14cm]{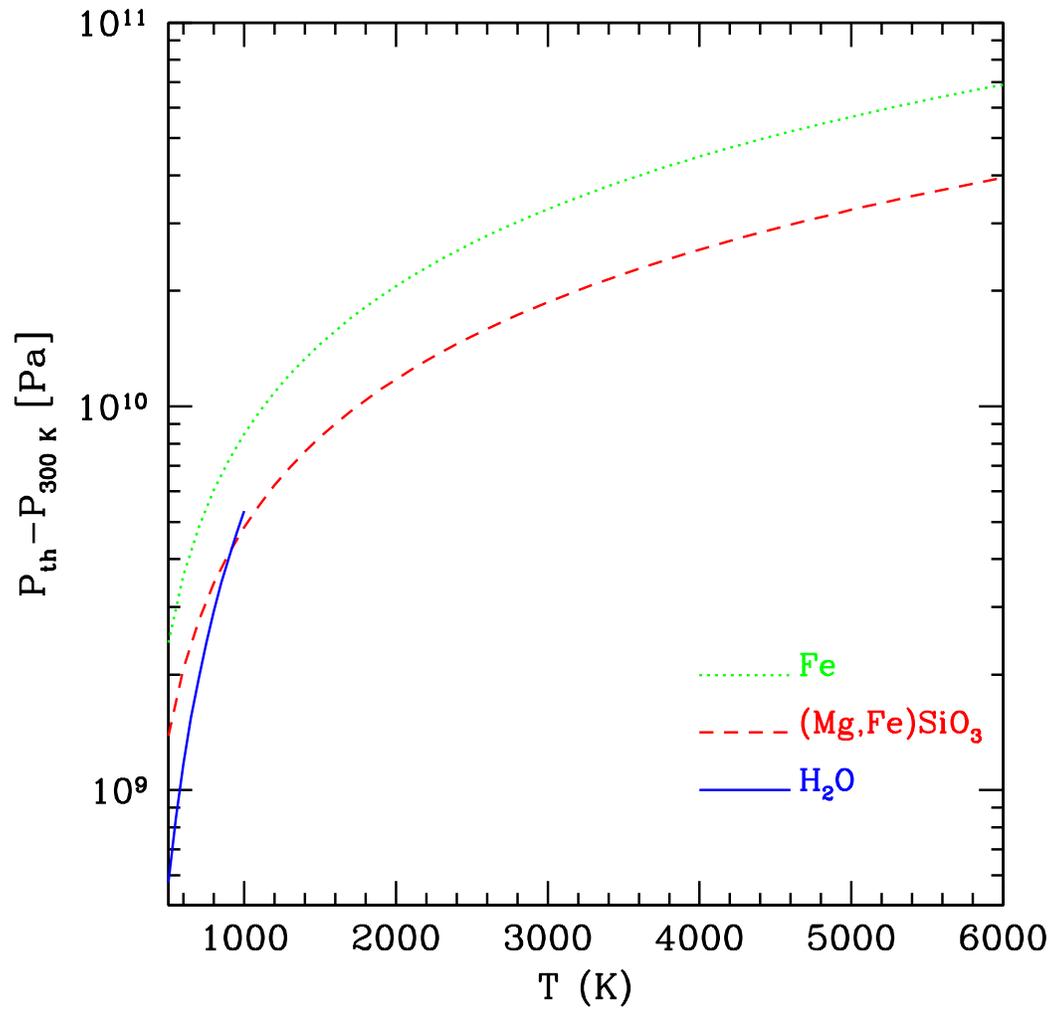}
\end{center}
\caption{Thermal pressure for Fe $(\varepsilon)$, 
(Mg,Fe)SiO$_3$, and H$_2$O. See text for details.
\label{fig:Pth}}
\end{figure}

\begin{figure}
\begin{center}
\includegraphics[width=14cm, height=14cm]{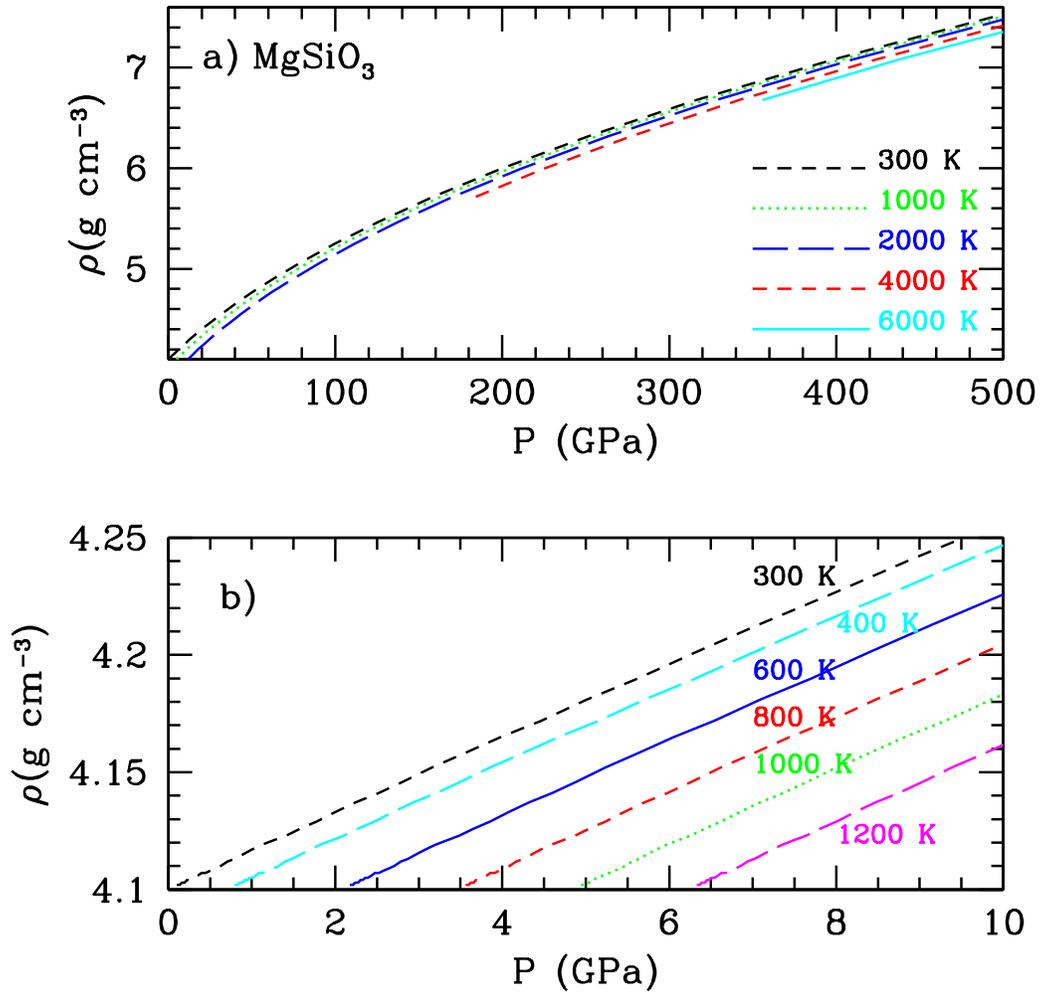}
\end{center}
\caption{Isotherms for MgSiO$_3$ (perovskite). Note that the
perovskite phase of silicate does not exist for the whole region of
low pressures shown in panel b, which is therefore shown to illustrate
the general properties of silicates.
\label{fig:Pthmg}}
\end{figure}

\begin{figure}
\begin{center}
\includegraphics[width=14cm, height=14cm]{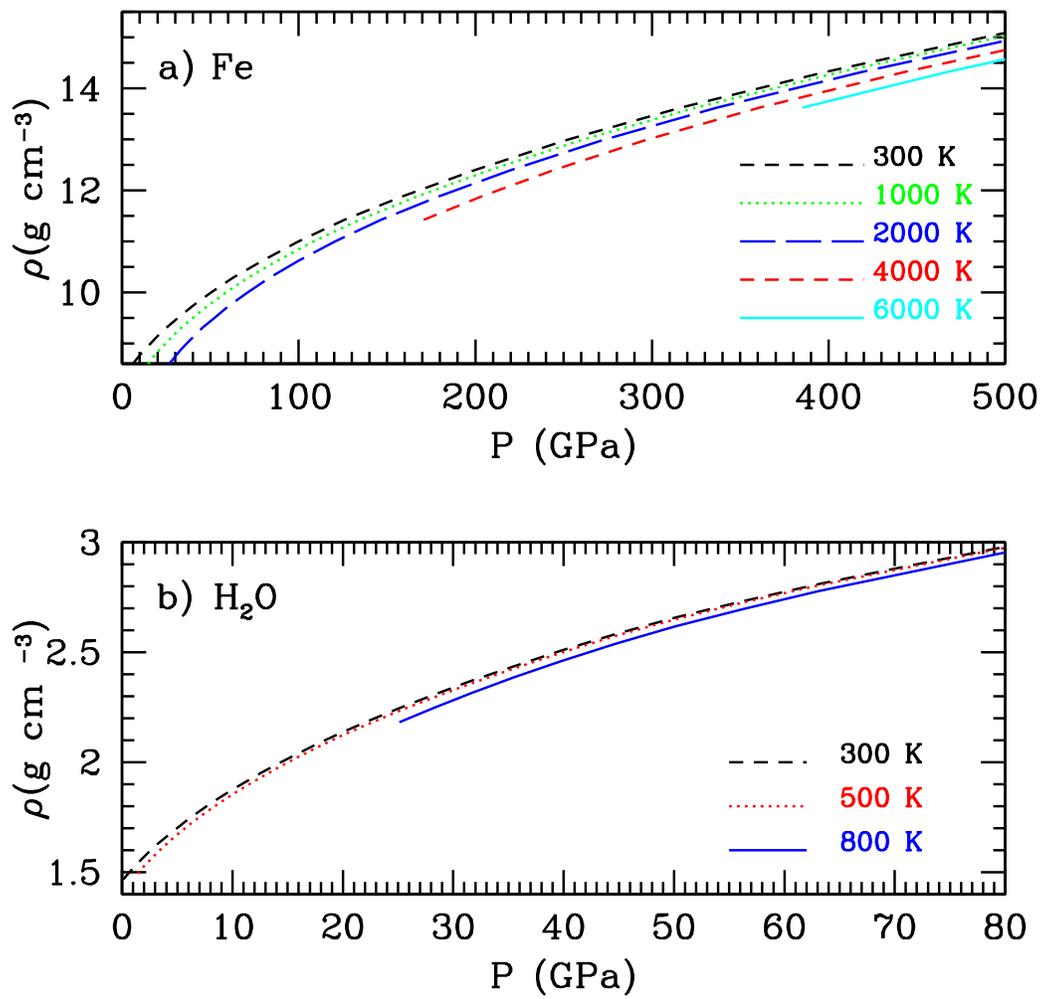}
\end{center}
\caption{Isotherms for Fe($\epsilon$) and H$_2$O ice
VII. See text for details.
\label{fig:Pthfeandh2o}}
\end{figure}

\begin{figure}
\begin{center}
\includegraphics[width=14cm, height=14cm]{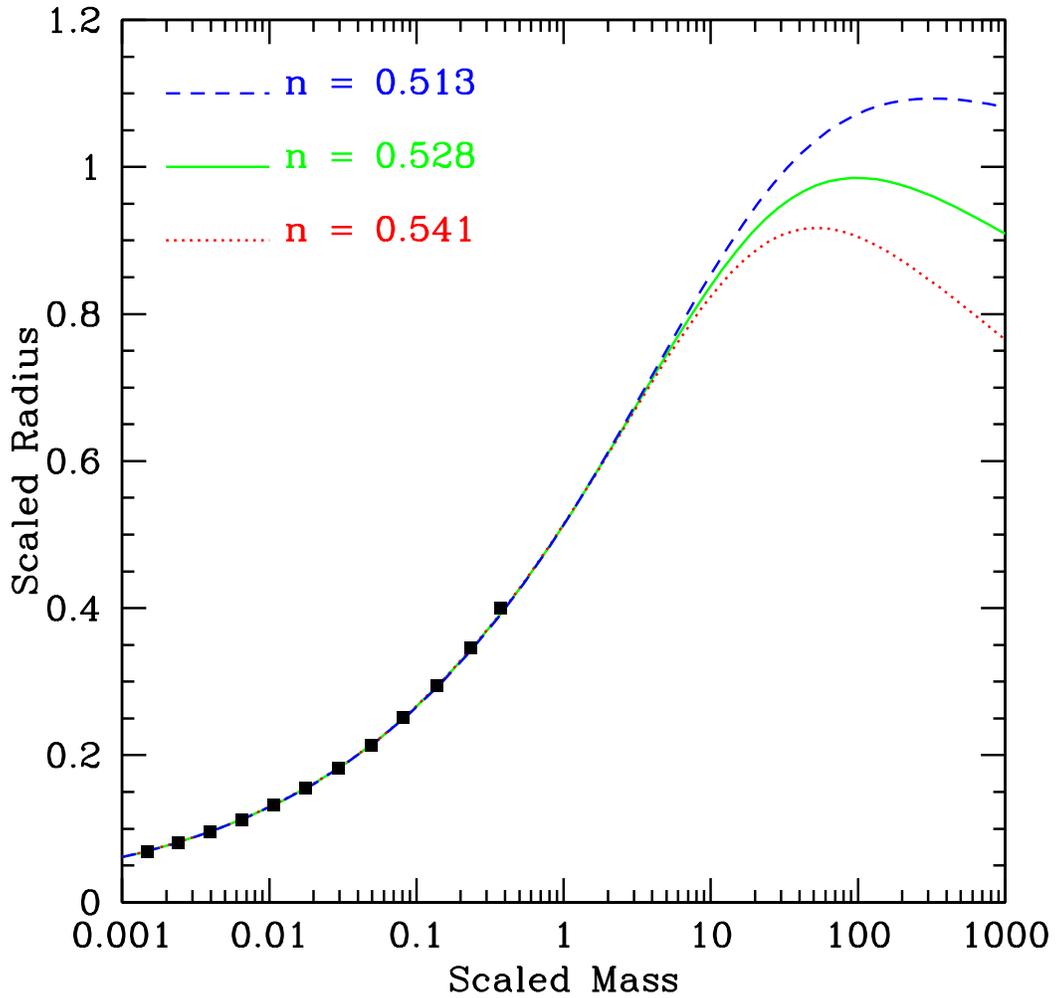}
\end{center}
\caption{Dimensionless mass-radius relationships for different
materials. The scaled mass and radius are given in
equations~(\ref{eq:scaledR}) and (\ref{eq:scaledM}). These
dimensionless mass-radius relationships depend only on the exponent,
$n$, in the modified polytrope EOSs ($n=0.513$ corresponds
to the H$_2$O modified polytrope, $n=0.528$ corresponds to Fe,
and $n=0.541$ corresponds to MgSiO$_3$).  For $M_s \lesssim 4$, the
mass-radius relationship for all materials takes approximately the
same functional form. The squares are the approximate analytical
scaled mass-radius relationship given in equation~(\ref{eq:msrs}),
which is valid for $M_s \ll 1$ and is the same for all $n$.
\label{fig:mrscaled} }
\end{figure}

\clearpage

\begin{figure}
\begin{center}
\includegraphics[width=14cm, height=14cm]{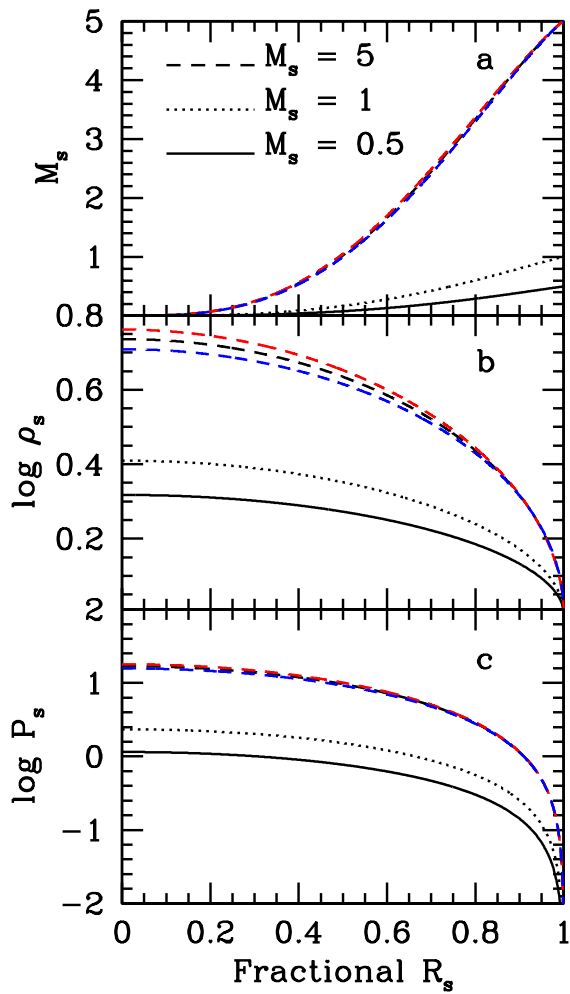}
\end{center}
\caption{Scaled interior structure models for solid exoplanets
computed from the polytropic-like EOSs.  Panel a: scaled mass
vs. fractional scaled radius. Panel b: scaled density vs. fractional
scaled radius. Panel c: scaled pressure vs. scaled fractional
radius. In all panels the solid, dotted, and dashed curves are for
scaled masses of 0.5, 1, and 5, respectively. Three different values
of $n$ are used 0.528 (Fe; black), 0.541 (MgSiO$_3$; red), and 0.514
(H$_2$O; blue). In all panels the solid and dotted curves are shown
for only a single value of $n$, because the curves for all $n$
overlap. The scaled interior structure solutions deviate from each
other for the $M_s= 5$ model.
\label{fig:rhoscaled} }
\end{figure}

\end{document}